\documentclass[showpacs,aps,prd,reprint,groupedaddress,superscriptaddress,preprintnumbers,floatfix,a4paper,nofootinbib,amsmath,amssymb]{revtex4-1}
\usepackage{color}
\usepackage{amsmath}
\usepackage{footmisc}
\usepackage{amssymb}
\usepackage{dcolumn}  
\usepackage{slashed}
\usepackage{isotope}
\usepackage{graphicx}
\usepackage{mathtools}
\usepackage[utf8]{inputenc}
\usepackage{multirow}
\usepackage{hyperref}
\usepackage{soul}
\hypersetup{colorlinks, linkcolor = [rgb]{0,0.0,0.75}, citecolor = [rgb]{0,0.0,0.75}, urlcolor = [rgb]{0,0.0,0.75}}

\makeatletter
\g@addto@macro\bfseries{\boldmath}
\makeatother

\begin{document}
\allowdisplaybreaks

\title{A study of doubly heavy tetraquarks in Lattice QCD}

\author{Parikshit Junnarkar}
\email{parikshit@theory.tifr.res.in}
\affiliation{Department of Theoretical Physics, \\ Tata Institute of Fundamental Research, 1 Homi Bhabha Road, Mumbai 400005. India.}

\author{Nilmani Mathur}
\email{nilmani@theory.tifr.res.in}
\affiliation{Department of Theoretical Physics, \\
Tata Institute of Fundamental Research, 1 Homi Bhabha Road, Mumbai 400005. India.}

\author{M Padmanath}
\affiliation{Instit\"ut  f\"ur  Theoretische Physik, Universit\"at Regensburg, D-93040 Regensburg, Germany.}



\begin{abstract}
We present the results of a lattice calculation of tetraquark states with quark contents $q_1q_2\bar{Q}\bar{Q}, \, q_1,q_2 \subset u,d,s,c$ and $Q \equiv b,c$ in both spin zero ($J=0$) and spin one ($J=1$) sectors.
This calculation is performed on three dynamical $N_f = 2 + 1 + 1$ highly improved staggered quark ensembles at lattice spacings of about 0.12, 0.09 and 0.06 fm.
We use the overlap action for light to charm quarks while a non-relativistic action with non-perturbatively improved coefficients with terms up to $\mathcal{O}(\alpha_s v^4)$ is employed for the bottom quark.
While considering charm or bottom quarks as heavy, we calculate the energy levels of various four-quark configurations with light quark masses ranging from the physical strange quark mass to that of the corresponding physical pion mass.
This enables us to explore the quark mass dependence of the extracted four-quark energy levels over a wide range of quark masses. 
The results of the spin one states show the presence of ground state energy levels which are below their respective thresholds for all the light flavor combinations.
Further, we identify a trend that the energy splittings, defined as the energy difference between the ground state energy levels and their respective thresholds, increase with decreasing the light quark masses and are maximum at the physical  point for all the spin one states. 
The rate of increase is however dependent on the light quark configuration of the particular spin one state. 
We also present a study of hadron mass relations involving tetraquarks, baryons and mesons arising in the limit of infinitely heavy quark and find that these relations are more compatible with the heavy quark limit in the bottom sector but deviate substantially in the charm sector.
The ground state spectra of the spin zero tetraquark states with various flavor combinations are seen to lie above their respective thresholds. 
\end{abstract}



\maketitle

\section{Introduction}
The past decade and a half has seen a remarkable number of discoveries in heavy hadrons. These new findings not only include regular mesons \cite{Dobbs:2008ec, Ablikim:2010rc, Adachi:2011ji, Lees:2011zp, Ablikim:2015dlj,Bhardwaj:2013rmw, Sirunyan:2018dff} and baryons \cite{Aaij:2017ueg, Aaij:2018yqz} but also involve exotic hadrons like tetra-\cite{Choi:2007wga, Aaij:2014jqa, 
Belle:2011aa} and pentaquarks \cite{Aaij:2015tga} while the structures of many are still puzzling (like many of the so called $X, Y$ and $Z$ states) \cite{Choi:2003ue, Aubert:2005rm, Yuan:2007sj, Voloshin:2007dx, Chen:2016qju, Hosaka:2016pey, Ali:2017jda, Aaij:2013zoa, Tanabashi:2018}. 
These hadrons, in particular, the multiquark states are reshaping our understanding of bound states and are providing new insights into the dynamics of strong interactions at multiple scales. 
Among the most notable multiquarks hadrons,  $Z_b(10630)$ and $Z^\prime_b(10650)$ were discovered first \cite{Belle:2011aa}, followed by $Z_c(4430)$ \cite{Choi:2007wga, Aaij:2014jqa, 
Belle:2011aa} and then $P_c$ pentaquarks \cite{Aaij:2015tga}.
Naturally these discoveries have kicked off a flurry of activities in heavy hadron physics, both theoretically and experimentally, and there is a real prospect of discovering more exotic hadrons, particularly with one or more bottom quark contents at various laboratories \cite{Karliner:2017qjm,Eichten:2017ffp,Francis:2016hui,Ali:2018xfq}.
The current status of these new discoveries, particularly on exotics are provided in various recent review articles \cite{Esposito:2016noz,Chen:2016qju, Hosaka:2016pey,Lebed:2016hpi,Olsen:2017bmm,Ali:2017jda}.

Theoretical studies of exotic hadrons are not new. Among the exotics, perhaps, tetraquarks are the most studied states. Historically, they were introduced by Jaffe \cite{Jaffe:1976ig} as color neutral states of diquarks and anti-diquarks{\footnote{A diquark can be interpreted as a compact colored object inside a hadron and is made out of two quarks (or antiquarks) in the $\overline{\mathbf{3}}(\mathbf{3})$ or $\mathbf{6}(\overline{\mathbf{6}})$ irrep of SU(3) and can have spin zero (scalar) or spin one (vector). With this model one can build rich phenomenology for mesons, baryons, as well as multiquark states.}}
in the context of describing light scalar mesons as tetraquarks and later for exotic spectroscopy \cite{Jaffe20051, Jaffe:2003sg}. 
Subsequently the diquark picture of tetraquarks was investigated in detail by many authors through various models \cite{Esposito:2016noz,Chen:2016qju,Hosaka:2016pey,Lebed:2016hpi,Ali:2017jda}.
Phenomenologically, a four-quark state can also be modelled as molecules \cite{Braaten:2003he,Guo:2017jvc}, hadroquarkonia \cite{Dubynskiy:2008mq,Braaten:2013boa} and also as threshold cusps \cite{Tornqvist:1995kr,Bugg:2008wu}, depending on how the four quarks interact mutually.

Though these models are effective with varying degree in describing these states, it is essential to have a first principles description of these strongly interacting hadrons. 
Lattice QCD, being a first principles non-perturbative method, ideally provides such an avenue to investigate these states comprehensively.  
The success of Lattice QCD, however, is still limited for these exotic states for multiple reasons. 
First, almost all such states that are observed, lie very close to their threshold energy levels. 
Though substantial progress has been made for resolving close-by states, it is essential to use novel techniques like distillation~\cite{Peardon:2009gh} that allows for the construction of large set of operators with the desired overlap onto the ground state which can then be computed using the variational principle \cite{Michael:1985ne,Luscher:1990ck}. 
Secondly, to identify a resonance state unambiguously from its non-interacting thresholds one has to perform the rigorous finite volume analysis \cite{Luscher:1990ck} of the discrete spectrum on multiple volumes and/or multiple momentum frames. 
Moreover these heavy hardons are very much susceptible to discretization error and a precise statement cannot be made unless one takes a controlled continuum limit of the results obtained at finite lattice spacings. All these issues, amount to a very large computationally intensive calculation which presumably will be carried out in future but currently is beyond the scope of any lattice group.

Current lattice QCD methods with available computational resources can however be an useful tool for studying hadrons which are far below their strong decay thresholds. For example, taking advantage of these methods and available computational resources one can study the deeply bound multiquark states to investigate whether such state exist in Nature. One can employ lattice methodology for a systematic search for these states using various spin and flavor combinations of interpolating operators and then dialing the quark masses, spanning over a wide range, can study the onset of a stable state with a large binding energy. 
In fact it has already been speculated several years ago that there may exist deeply bound tetraquark states in the heavy quark limit. 
Using one pion exchange between the ground state $Q\bar{q}$ mesons, Manohar and Wise showed that QCD contains stable (under strong interactions) four-quark $QQ\bar{q}\bar{q}$ hadronic states in the infinite quark mass limit, and for the bottom quark this binding could well be sufficiently large \cite{Manohar:1992nd}.

The heavy tetraquarks are also studied recently using heavy quark effective theory  \cite{Mehen:2017nrh,Eichten:2017ffp}, quark models \cite{Maiani:2004vq, Cui:2006mp,Ebert:2007rn,Ebert:2008kb, Ali:2014dva,Karliner:2017qjm,Richard:2018yrm, Wilcox:2018tau}, QCD sum rules \cite{Wang:2010uf,Wang:2013vex,Kleiv:2013dta} and large $N_c$ calculations \cite{Weinberg:2013cfa,Knecht:2013yqa,Cohen:2014tga}{\footnote{There are many model calculations on tetraquarks and for a detail reference list readers may want to see review articles \cite{Esposito:2016noz,Chen:2016qju, Hosaka:2016pey,Lebed:2016hpi,Olsen:2017bmm,Ali:2017jda}}}. The proposed doubly bottom tetraquark state and its isospin cousins are believed to be strong interaction stable states with relatively long life times.
Recently lattice QCD calculations \cite{Francis:2016hui, Junnarkar:2017sey} and a lattice-QCD-potential based study \cite{Bicudo:2016ooe,Bicudo:2015vta,Bicudo:2015kna} also identified  a particular exotic flavor-spin combination of two bottom quarks, namely $ud \bar{b} \bar{b}$, with a prediction of a deeply bound state which lies below its non-interacting two-meson threshold. It is thus quite crucial to investigate such and similar states using a detailed lattice QCD study by incorporating various heavy and light flavor combinations along with different spin combinations and at multiple lattice spacings.

In this work we carry out such a calculation where we use both the charm and bottom as heavy quarks, and then vary the light quark masses from the strange quark mass to the corresponding lower pion masses leading to various tetraquark states: $q_1q_2\bar{Q}\bar{Q}, q_1,q_2 \subset u,d,s,c$ and $Q \equiv b,c$ with both spin zero ($J=0$) and spin one ($J=1$). 
These are computed at three lattice spacings of $\sim$  0.12, 0.09 and 0.06 fm, to investigate the discretization effects on these heavy hadrons.
We use the relativistic overlap action, for light to charm quarks while a non-relativistic action with non-perturbatively improved coefficients with terms up to $\mathcal{O}(\alpha_s v^4)$ is employed for the bottom quark. Our results for the spin one tetraquarks indicate the presence of energy levels below the respective thresholds for all light flavor combinations with doubly heavy, in particular, for doubly bottom quarks.
The results for spin zero tetraquarks, which are the flavor symmetric cousin states of the spin one counterparts, however indicate the respective energy levels  are above their lowest strong decay two-meson thresholds.
In addition to computing the ground state spectra, we also present a lattice study of the hadron mass relations between tetraquarks, heavy baryons and mesons arising from the heavy quark symmetry. 
In future we will incorporate also the finite volume study so that more quantitative conclusions about the pole structures of these tetraquark states can be made, particularly for the near-threshold states.

The paper is organized as follows: In section~\ref{sec:lattice_setup} we elaborate
the lattice set up, actions employed and the  quark mass combinations that we use for this work.
Section~\ref{sec:operators} provides details of the tetraquark operators and the flavor-spin combinations that we employ in this work.  In section~\ref{sec:results}, with the details of analysis method we present our results, first for the spin one sector followed by the spin zero sector. Finite volume effects on our results are discussed thereafter. A discussion on the hadron mass relations with the heavy quark symmetry is followed afterwards. 
Finally conclusions from this work are discussed in section~\ref{sec:conclusions}.


\section{\label{sec:lattice_setup} Lattice setup}
We perform this calculation on three dynamical 2+1+1 flavors lattice ensembles 
generated by the MILC Collaboration~\cite{Bazavov:2012xda}.
These  ensembles, with lattice sizes  $24^3 \times 64$, $32^3 \times 96$ and $48^3 \times 144$, at gauge couplings $10/g^2 = 6.00, 6.30$ and $6.72$, respectively, were generated with the HISQ action and with the one-loop, tadpole improved Symanzik gauge action with coefficients corrected through $\mathcal{O}(\alpha_{s}a^2,n_f\alpha_sa^2)$ \cite{Follana:2006rc}. 
The masses of strange and charm quarks on these ensembles are set to their physical values while the light sea quark masses are set such that $m_s/m_l = 5$. The lattice spacings as measured using the $r_1$
parameter for the set of ensembles used here are 0.1207(11), 0.0888(8) and 0.0582(5) fm, respectively \cite{Bazavov:2012xda}.
Further details of these lattice QCD ensembles can be found in Ref. \cite{Bazavov:2012xda}.
\begingroup
\renewcommand*{\arraystretch}{1.2}
\begin{table}[ht]
	\centering
	\begin{tabular}{cccc}\hline \hline
		$N^3_s \times N_t$  & a(fm) & $am_q$ & $m_\pi$ (MeV)  \\ \hline \hline
		$24^3 \times 64$    & 0.1207(14)    & 0.0738 & 689 \\ 
									&                            & 0.054 &  589 \\ 
						        	&                            & 0.045 &  539  \\ 
	 							 	&                            & 0.038 &  497  \\
	 							    &                            & 0.030 &  449  \\
	 							    &                            & 0.024 &  400  \\
	 							    &                            & 0.020 &  367  \\
	 							    &                            & 0.0165 & 337  \\
	 							    &							  & 0.0125 & 297  \\
	 							    &                            & 0.0090 & 257  \\
	 							    &                            & 0.0075 & 237  \\
	 							    &                            & 0.0060 & 216  \\
	 							    &                            & 0.0051 & 202  \\
	 							    &                            & 0.0042 & 186  \\
	 							    &                            & 0.0028 & 153  \\
	 							    		\hline \hline
		$32^3 \times 96$ & 0.0888(5)& 0.049 & 688   \\ 
									&                           & 0.030 & 537   \\
									&                           & 0.020 & 441  \\
									&                           & 0.016  & 396  \\
									&                           & 0.0135 & 367  \\
									&                           & 0.012  & 345  \\
		\hline \hline
		$48^3 \times 144$ & 0.0582(5)& 0.028 & 685  \\
									 &                          & 0.025 & 645  \\
									 &                          & 0.020 & 576  \\
									 &                          & 0.018 & 545  \\
		 \hline \hline	\end{tabular}
	\caption{\label{tab:pars}{Parameters of ensembles used in this work}}
\end{table} 
\endgroup

In the valence sector, for light, strange and charm quarks,
we employ the overlap fermion action \cite{Neuberger:1997fp,Neuberger:1998wv}, which has exact chiral symmetry at finite lattice spacings~\cite{Neuberger:1997fp, Neuberger:1998wv, Luscher:1998pqa}
and is automatically $\mathcal{O}(ma)$ improved.
The numerical implementation of the overlap fermion is carried out following the methods in Refs. \cite{Chen:2003im, Li:2010pw}. A wall source smearing is utilized to calculate the light to charm quark overlap propagators on Coulomb gauge fixed lattices.
In Table I, we list the quark masses and corresponding pion masses that we use for this calculation. The strange quark mass is tuned by equating the lattice estimate of the $\bar{s} s$ pseudoscalar meson mass to 688.5 MeV \cite{Chakraborty:2014aca,Basak:2012py,Basak:2013oya}. 
We follow the Fermilab prescription of heavy quarks for tuning the charm quark mass~\cite{ElKhadra:1996mp}. We tune it by equating the spin-averaged kinetic mass of the $1S$ charmonia 
($a\bar{M}_{kin}(1S) = {3\over 4} aM_{kin}(J/\psi) + {1\over 4} aM_{kin}(\eta_c)$) to its experimental value, 3068.6 MeV~\cite{Tanabashi:2018}. The tuned bare charm quark masses are found to be 0.528, 0.427 and 0.290
on coarse to fine lattices respectively, all of which satisfy $m_ca << 1$  ensuring reduced discretization artifacts in this calculation. Details on the charm quark mass tuning can be found in Refs. \cite{Basak:2012py,Basak:2013oya}.

For the bottom quarks, we employ a non-relativistic QCD (NRQCD) formulation~\cite{Lepage:1992tx}. In the NRQCD Hamiltonian we include all the terms up to $1/{M_0^2}$ as well as the leading term of the order of $1/{M_0^3}$, where $M_0 = am_b$ is the bare mass of the bottom quarks in lattice units \cite{Lewis:2008fu}.
The bottom quark propagators are obtained by the usual time evolution of the NRQCD Hamiltonian, $H = H_0 + \Delta H$, where  the interaction term, $\Delta H$, is given by,
\begin{eqnarray}
\Delta H &=& -c_1 \frac{(\Delta^{(2)})^2}{8(am_b)^3} + c_2 \frac{i}{8 (am_b)^3}(\nabla \cdot \tilde{E}  - \tilde{E} \cdot \nabla) \nonumber \\
&& -c_3 \frac{1}{8 (m_b)^2} \sigma \cdot (\nabla \times \tilde{E}  - \tilde{E} \times \nabla) - c_4 \frac{1}{2 am_b} \sigma \cdot \tilde{B} \nonumber \\
&& +  c_5 \frac{(\Delta^{(4)}}{24 a m_b} - c_6 \frac{(\Delta^{(2)})^2}{16 (am_b)^2}. \\ \nonumber
\end{eqnarray}
Here $c_1..c_6$ are the improvement coefficients, and for the fine lattice we use their tree level values while for coarser two lattices we employ their non-perturbative values as estimated by the HPQCD collaboration~\cite{Dowdall:2011wh} on the same set of lattices.
To tune the bottom quark mass we first calculate the kinetic mass of the spin average $1S$ bottomonia,
\begin{equation}
a{M}_{Kin} = \frac{3}{4} a M_{Kin}(\Upsilon) + \frac{1}{4} a M_{Kin}(\eta_b),
\end{equation}
from the relativistic energy-momentum dispersion relation $aM_{Kin} = ((ap)^2  - (a \Delta E)^2)/(2a \Delta E)$, and then equate it with its experimental value. Details on the bottom quark mass tuning is given in Ref. \cite{Mathur:2016hsm}.

With this setup of light, strange, charm and bottom quark propagators, we proceed to calculate the tetraquark correlators from the interpolating fields with various flavor-spin combinations that we discuss in the next section.

\section{\label{sec:operators}Four-quark interpolating operators}
In this section, we describe four-quark interpolating fields (operators) that we
employ in this work.  We construct these operators with two heavy and two light quarks 
 and with the total spin $J=0$ and 1. As in  Ref.~\cite{Francis:2016hui}, for both spins we construct
two type of operators, with a goal that one overlaps onto a tetraquark state
of given quantum numbers and the other one overlaps onto the lowest strong decay two-meson states of the same quantum numbers.
The {\it tetraquark-type} operators are constructed using the diquark
prescription of Jaffe \cite{Jaffe20051,Jaffe:2003sg} where a color neutral hadronic operator is constructed as a product of diquarks and anti-diquarks.
These diquarks (anti-diquarks) can be  in the $\overline{\mathbf{3}}_c (\mathbf{3}_c)$ or $\mathbf{6}_c(\overline{\mathbf{6}}_c)$ of the color SU(3) irreducible representation (irreps). 
Phenomenologically, the one gluon exchange model \cite{Jaffe20051,Jaffe:2003sg} favors an attractive interaction of two quarks and is in the $\bar{\mathbf{3}}$ irrep of SU(3).
In this work, we construct tetraquark operators with both irreps of SU(3).

In the spin $J=1$ sector, we use diquarks and anti-diquarks with the following configuration:
\begin{equation} 
(l_1,l_2) \rightarrow (\overline{\mathbf{3}}_c,0,F_A), \quad \quad (\bar{Q}, \bar{Q}) \rightarrow (\mathbf{3}_c,1,F_s).
\end{equation} 
The light quark $(l_1,l_2; l_1\neq l_2)$ combinations are constructed with  color, spin and flavor degrees of freedom as antisymmetric and are restricted within $\subset (u,d,s,c)$.
The heavy quark combination $(\bar{Q},\bar{Q})$ is constructed with color antisymmetric $\mathbf{3}_c$, forced by $(l_1,l_2)$ being in the $\bar{\mathbf{3}}_c$, and since flavor is manifestly symmetric the spin is also symmetric.
This combination is restricted to only heavy flavors $ \subset (\bar{c},\bar{b})$ with further restriction of  $Q \neq l_1 \neq l_2 $.
With these diquarks and anti-diquarks, a spin one {\it tetraquark-type} operator of flavor $(l_1 l_2 \bar{Q} \bar{Q})$  is constructed as :
\begin{equation}
\label{Eq:S1T}
\mathcal{T}^{\mathbf{1}}(x) = (l_1)^a_{\alpha}(x) \  (C \gamma_5)_{\alpha \beta} \ (l_2)^{b}_{\beta}(x) \ \bar{Q}^a_{\kappa}(x) (C \gamma_i)_{\kappa \rho} \ \bar{Q}^b_{\rho}(x) 
\end{equation}
 The label  $x$ is a shorthand notation for $(\vec{x}, t)$ where $\vec{x}$ is the spatial local site and $t$ is the timeslice.
We then construct the {\it two-meson-type} operators corresponding to each flavor of $(l_1 l_2 \bar{Q} \bar{Q})$ tetraquark operator, $\mathcal{T}^{\mathbf{1}}(x)$, with the appropriate flavor antisymmetry as: 
\begin{eqnarray}
\label{Eq:2M1}
\mathcal{M}^{\mathbf{1}}(x) &=& M_1(x)M_2^{*}(x) - M_2(x) M_1^{*}(x) \nonumber \\
M_{1,2}(x) &=& (l_{1,2})^a_\alpha(x) \ (\gamma_5)_{\alpha \beta} \ \bar{Q}^{a}_{\beta}(x) \nonumber \\  M^*_{1,2}(x) &=& (l_{1,2})^a_\alpha(x) \ (\gamma_i)_{\alpha \beta} \ \bar{Q}^{a}_{\beta}(x). 
\end{eqnarray}
The tetraquark operator $\mathcal{T}^{\mathbf{1}}(x)$ is related to the two-meson product $M_1(x) M^*_2(x)$ via a Fierz transformation and the relation is explicitly shown in the appendix of Ref.~\cite{Padmanath:2015era} with the appropriate change in flavor labels.
The various flavor and isospin ($I$) combinations that we explore for these spin one {\it tetraquark-type} and {\it two-meson-type} operators are tabulated in Table \ref{tab:spinone}. 
\begingroup
\renewcommand*{\arraystretch}{1.9}
\begin{table}[ht]
	\centering
	\caption{\label{tab:spinone}{The {\it tetraquark-type} and {\it two-meson-type} operators that we study in this work with possible flavor combinations and allowed isospin ($I$) in the spin one sector. The last column shows the range of pion masses that we use for the light quarks on the coarsest lattice spacing.}}
	\begin{tabular}{c c c | c}\hline \hline 
	 $(l_1 l_2 \bar{Q} \bar{Q})$ &  $[(M_1 M^*_2) (M_2 M^*_1)]$ & $I$ & $m_\pi$ (MeV) \\ \hline \hline
	 $ud\bar{b}\bar{b}$  & $(B  B^{0*}) (B^0 B^*)$ & 0 & (257 - 688)\\ \hline
	 $us\bar{b}\bar{b}$  & $(B B^{*}_s) (B_s B^*)$ & $\frac{1}{2}$ & (186 - 688)  \\ \hline
	 $uc\bar{b}\bar{b}$  & $(B B^{*}_c) (B_c B^*)$ & $\frac{1}{2}$ & (153 - 688) \\ \hline
	 $ud\bar{c}\bar{c}$  & $(D D^{0*}) (D^0 D^*)$ & 0  & (257 - 688)\\ \hline
	 $us\bar{c}\bar{c}$  & $(D D^{*}_s) (D_s D^*)$ & $\frac{1}{2}$ & (257 - 688)  \\ \hline \hline
	\end{tabular}
\end{table}
\endgroup

For the spin zero sector, we employ following diquark anti-diquark configuration where both diquarks are with spin zero: 
\begin{equation} 
(l,l) \rightarrow (\mathbf{6}_c,0,F_S), \quad \quad (\bar{Q}\bar{Q}) \rightarrow (\overline{\mathbf{6}}_c,0,F_s).
\end{equation} 
The combination $(l,l)$ being manifestly flavor symmetric requires the color degree of freedom to be in the $\mathbf{6}_c$. 
For the combination $(\bar{Q},\bar{Q})$, the color degree of freedom is consequently restricted to $\overline{\mathbf{6}}_c$  while the flavor degree of freedom is manifestly symmetric.  In the above expression, for the combination $(l,l)$ we incorporate the flavors $(u,s,c)$ while both $c$ and $b$ are used for $Q$.
A spin zero {\it tetraquark-type} operator of flavor $(ll \bar{Q}\bar{Q}) $ constructed from the product of the aforementioned diquarks and anti-diquarks is given by: 
\begin{equation}
\label{Eq:S0T}
\mathcal{T}^{\mathbf{0}}(x) = l^a_\alpha(x) (C \gamma_5)_{\alpha \beta} l^b_\beta(x) \ \bar{Q}^b_\kappa(x) (C \gamma_5)_{\kappa \rho} \bar{Q}^a_\rho(x).
\end{equation}
As previously, we also construct a {\it two-meson-type} operator with the same quantum number of that of $(ll \bar{Q} \bar{Q})$ 
 and is given by: 
\begin{equation}
\label{Eq:2M0}
\mathcal{M}^{\mathbf{0}}(x) = \bar{Q}^a_\alpha(x) (\gamma_5)_{\alpha \beta} l^a_\beta(x) \  \bar{Q}^b_\kappa(x) (\gamma_5)_{\kappa \rho} l^b_\rho(x). \\
\end{equation}
In Table~\ref{tab:spinzero} we tabulate the spin zero tetraquark configurations with the possible flavour combinations with the above flavour-spin configurations.
\begingroup
\renewcommand*{\arraystretch}{1.9}
\begin{table}[ht]
	\centering
	\caption{\label{tab:spinzero}{The {\it tetraquark-type} and {\it two-meson-type} operators for various flavors of in the spin zero sector. The range of pion masses used for $uu\bar{b}\bar{b}$ and $uu\bar{c}\bar{c}$ states is indicated in the last column. All other states computed at their physical quark mass.}}
	\begin{tabular}{c c c | c}\hline \hline 
	 $(l_1 l_2 \bar{Q} \bar{Q})$ &  $(M_1M_2)$ & $I$ & $m_{\pi}$ (MeV)\\ \hline \hline
	 $uu\bar{b}\bar{b}$  & $(BB)$ & 1 & (337 - 688)\\ \hline
     $uu\bar{c}\bar{c}$  & $(DD)$ & 1  & (297 - 688)\\ \hline
	 $ss\bar{b}\bar{b}$  & $(B_sB_s)$ & 0 & -\\ \hline
	 $cc\bar{b}\bar{b}$  & $(B_cB_c)$ & 0 & -\\ \hline
	 $ss\bar{c}\bar{c}$  & $(D_sD_s)$ & 0 & -\\ \hline \hline
	\end{tabular}
\end{table}
\endgroup

With the operators so constructed, we proceed to compute the correlation matrices of all the possible combinations of these operators for a given spin and flavor, and then extract the associated energy states from the generalized eigenvalue solutions. In the next section we discuss this in detailed.

\section{Results}\label{sec:results}

In this section, first we elaborate the analysis procedure that we utilize to extract the energy levels from the matrix of correlation functions constructed from the interpolating fields mentioned above. Results obtained will be discussed after that.
\subsection{Analysis Methods}
To evaluate the energy levels corresponding to the operators discussed in ~\ref{sec:operators}, we first construct a correlator matrix of these operators and then use the variational method \cite{Michael:1985ne,Luscher:1990ck}. 
This matrix of correlation functions $C_{ij}(t)$ is given as:
\begin{equation}
C_{ij}(t) = \sum_{\vec{x}}  \langle 0 | \mathcal{O}_i(\vec{x},t) \mathcal{O}_j^\dagger(\vec{0},0) | 0 \rangle , 
\end{equation}
where the operator $\mathcal{O}_i(\vec{x},t) \in \big\{\mathcal{T}^{k}(\vec{x},t),\mathcal{M}^{k}(\vec{x},t)\big\}$ is either a {\it tetraquark-type} operator or a {\it two-meson-type} operator of a particular spin $k$.
 For the spin one tetraquark states $\mathcal{O}_i$'s correspond to Eqs.~(\ref{Eq:S1T}) and (\ref{Eq:2M1}) whereas for the spin zero states these are from Eqs.~(\ref{Eq:S0T}) and (\ref{Eq:2M0}). We analyze each spin sector separately.
After constructing the correlation matrix, $\mathbf{C}(t)$, for a given spin and flavor combination, we solve a generalized eigenvalue problem (GEVP) to obtain the two energy levels \cite{Michael:1985ne,Luscher:1990ck}. 
The standard methods for GEVP~\cite{Michael:1985ne,Luscher:1990ck,Blossier:2009kd,Green:2014dea} are typically suited for a Hermitian correlator matrix.
We note that since we are using a wall source, the  correlator matrix is non-Hermitian\footnote{The same correlator matrix is found to be hermitian when computed with unsmeared point sources and sink.}. Hence we employ a variation of GEVP method, named as eigenvector method, involving eigenvector projection in evaluating the ground state energies ~\cite{Francis:2018qch}.
The method involves using the left and right eigenvectors of the correlator matrix to construct the principal correlator as discussed below: 
\begin{enumerate}
\item Compute left and right eigenvectors of the correlator matrix $\mathbf{C}(t)$ at chosen time-slices $(t_1, t_0)$ as:
\begin{eqnarray}
\mathbf{C}(t_1) v_{R,n}(t_1,t_0) &=&  \lambda_n(t_1,t_0) \mathbf{C}(t_0) v_{R,n}(t_1, t_0) \nonumber \\
 v_{L,n}(t_1,t_0)  \mathbf{C}(t_1) &=&  \lambda_n(t_1,t_0) v_{L,n}(t_1, t_0) \mathbf{C}(t_0).
\end{eqnarray}
The time-slices $(t_1, t_0)$ are chosen such that $t_1/t_0 > 2$ and $t_1$ chosen in the region where the correlator is expected to be dominated by the ground state.   
\item The eigenvectors $v_{L,R,n}(t_1,t_0)$ are then used to construct the principal correlator as: 
\begin{equation} \label{Eq:approach_3}
\Lambda_n(t) = v^\dagger_{L,n} (t,t_0) \mathbf{C}(t) v_{R,n}(t_1, t_0) ,
\end{equation}
and the effective masses are then obtained from $m_{n,\text{eff}} = \text{log}(\Lambda_n(t)/\Lambda_n(t+ \delta t)).$
\end{enumerate}
For a Hermitian correlator matrix, the left and right eigenvectors will be identical and hence this method will be the same as standard methods~\cite{Michael:1985ne,Luscher:1990ck,Blossier:2009kd,Green:2014dea}.
For a non-Hermitian correlator, the source and sink operators are accordingly rotated by the left and right eigenvectors respectively. To check the effects of non-hermiticity we also solve GEVP with the standard methods ~\cite{Michael:1985ne,Luscher:1990ck,Blossier:2009kd,Green:2014dea}. 
We find consistent results with our preferred eigenvector method and the results from the eigenvector method being more stable.

The  principal correlators thus obtained correspond to two energy levels and the ground state energy is computed from the lowest one. 
On the other hand, we calculate the non-interacting two meson threshold from the sum of the ground state masses of the two mesons involved. 
We then compare the lowest energy level obtained from GEVP solution with the non-interacting two-meson threshold and evaluate the energy splitting between them  as:
\begin{equation}
\label{Eq:eff_split}
\Delta E^k  = E_{\mathcal{T}^k} - E_{2M},
\end{equation}
where $E_{\mathcal{T}^k}$ is the ground state energy obtained from the principal correlator of GEVP while $E_{2M} = E_{M_1} + E_{M_2} $ is the energy of the non-interacting two meson ($M_1$ and $M_2$) threshold.
The above energy splitting ($\Delta E^k$) can be evaluated directly by fitting the two data sets separately and then computing the difference on each resample.
Alternatively, this can also be evaluated by taking the Jackknife ratio of the principal correlator ($\Lambda(t)$) of GEVP to two-meson correlators, $M_1(t) \times M_2(t)$) as: 
\begin{equation}\label{Eq:approach_4}
\Lambda^\prime (t) = \frac{\Lambda(t)}{M_1(t) \times M_2(t)} \rightarrow \mathcal{A} e^{-\Delta E^k t} + ...
\end{equation}
A fit to the ratio correlator ($\Lambda^\prime(t)$) will then yield directly the energy splitting with respect to the relevant threshold.
Such a construction offers the advantage of reducing the systematic errors through Jackknifing. 
However in using such an effective correlator, caution must be exercised as this construction can produce spurious effects since the saturation of the ground states of the numerator and the denominator may not happen at the similar time slices.
In this work, in estimating the energy splitting, we utilize both the direct and ratio methods and find consistent results. 
However, as expected we find smaller uncertainties in the ratio method.
 \begin{figure}
	\includegraphics[width=1.0\linewidth]{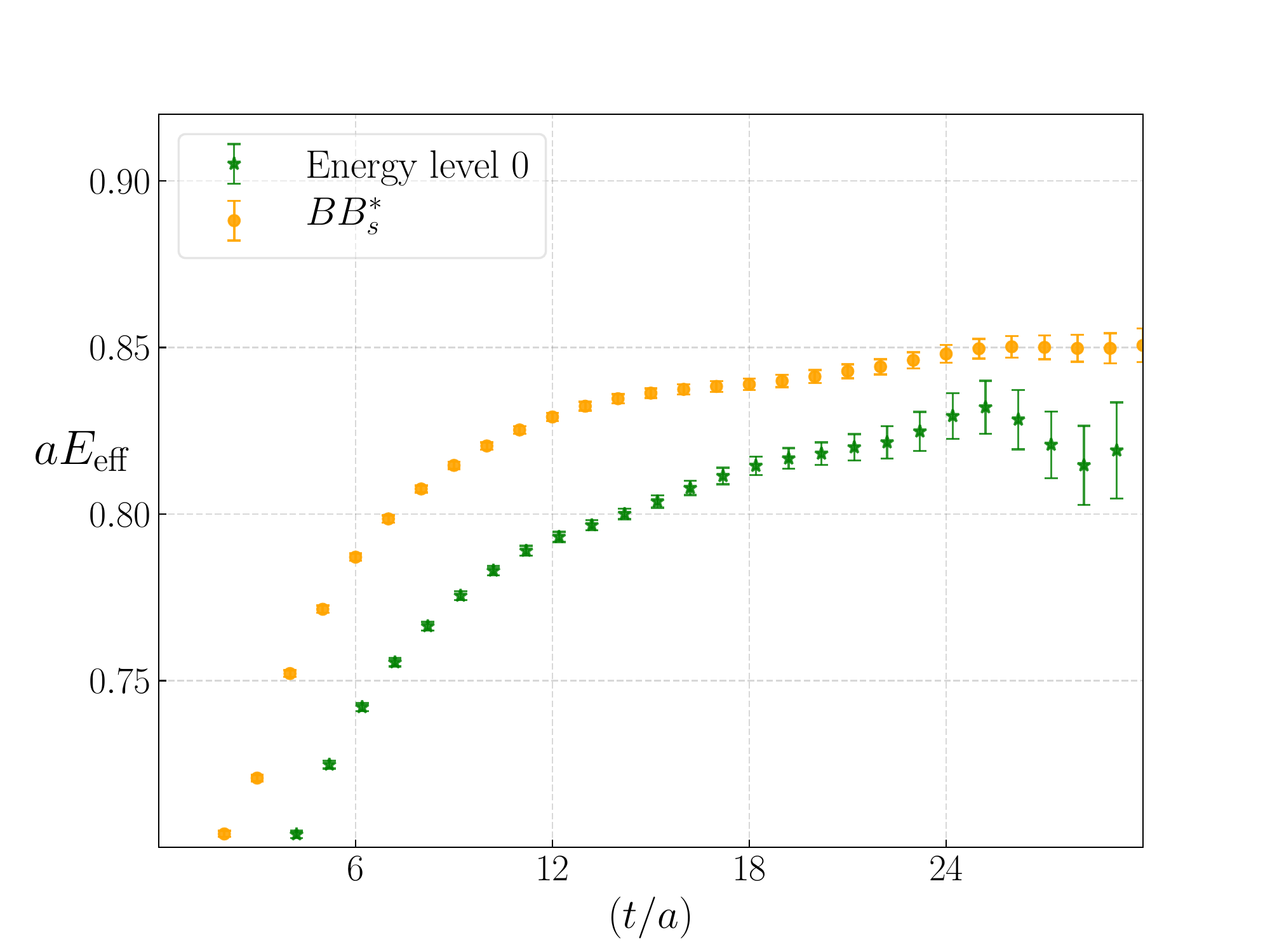}
	\caption{\label{fig:effmass} Effective mass of the ground state energy level (data in green) obtained from GEVP solution for the spin 1, $us\bar{b}\bar{b}$ tetraquark state at $m_\pi = 688$ MeV and $a=0.0582$ fm. The data in orange is the effective mass of the threshold correlator $B B^*_s$.}
\end{figure}
We now present the results obtained through above mentioned analysis.

\subsection{Spin one tetraquarks $J^P=1^+$\label{subsec:spinone}}
\begin{figure*}[ht]
 \includegraphics[width=1.0\linewidth]{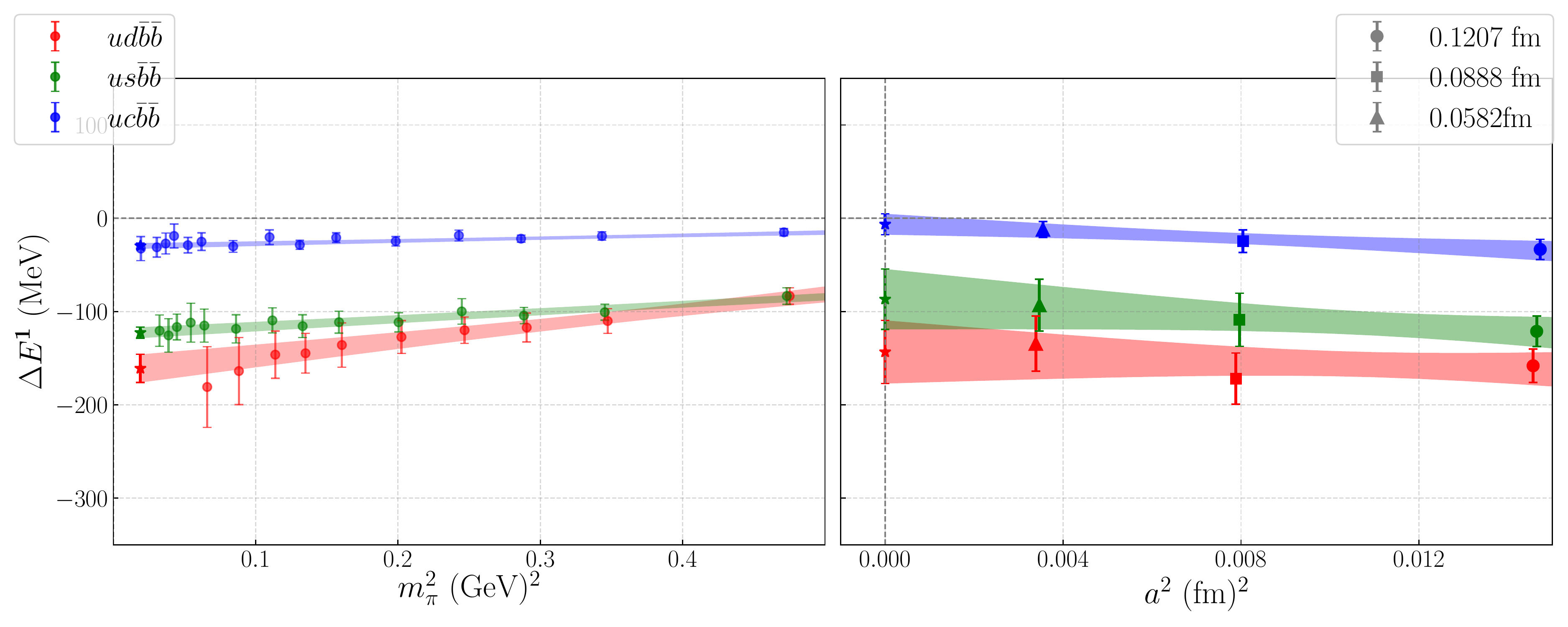}
 \caption{\label{fig:uqbb_1} Results of $ud\bar{b}\bar{b}, us\bar{b}\bar{b}$ and $uc\bar{b}\bar{b}$ doubly bottom tetraquark states color coded in red, green and blue respectively in both panels.  Left panel: Energy splittings at several pion masses at $a = 0.1207$ fm for each of the states. The fit bands indicate a chiral extrapolation fit as per Eq.~(\ref{Eq:chiral}) color coded appropriately for each state. Right panel: Continuum extrapolation results as per Eq.~(\ref{Eq:conti}) from three lattice spacings. The data point at each lattice spacing is the result of the chiral extrapolation to the physical pion mass at that lattice spacing.}
 \end{figure*}
 \begin{figure*}[ht]
   \includegraphics[width=1.0\linewidth]{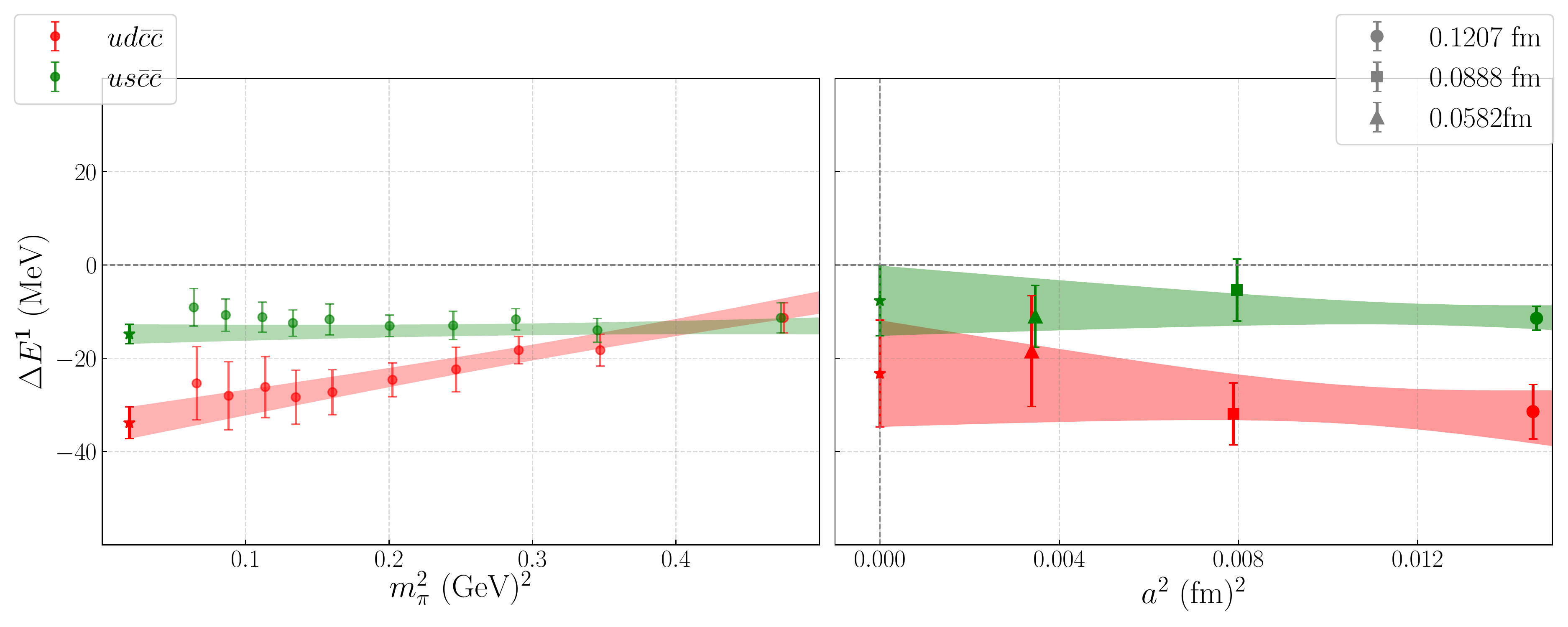}
   \caption{\label{fig:uqcc} Results of $ud\bar{c}\bar{c}$ and $us\bar{c}\bar{c}$ doubly charm tetraquark states color coded in red and green in both panels.  Left panel: Effective splittings at several pion masses at $a = 0.1207$ fm for each of the states. The fit bands indicate a chiral extrapolation fit as per Eq.~(\ref{Eq:chiral}) color coded appropriately for each state. Right panel: Continuum extrapolation results as per Eq.~(\ref{Eq:conti}) from three lattice spacings. The data point at each lattice spacing is the result of the chiral extrapolation to the physical pion mass at that lattice spacing.}
  \end{figure*}

\begin{figure}[h]
	\includegraphics[width=1.1\linewidth]{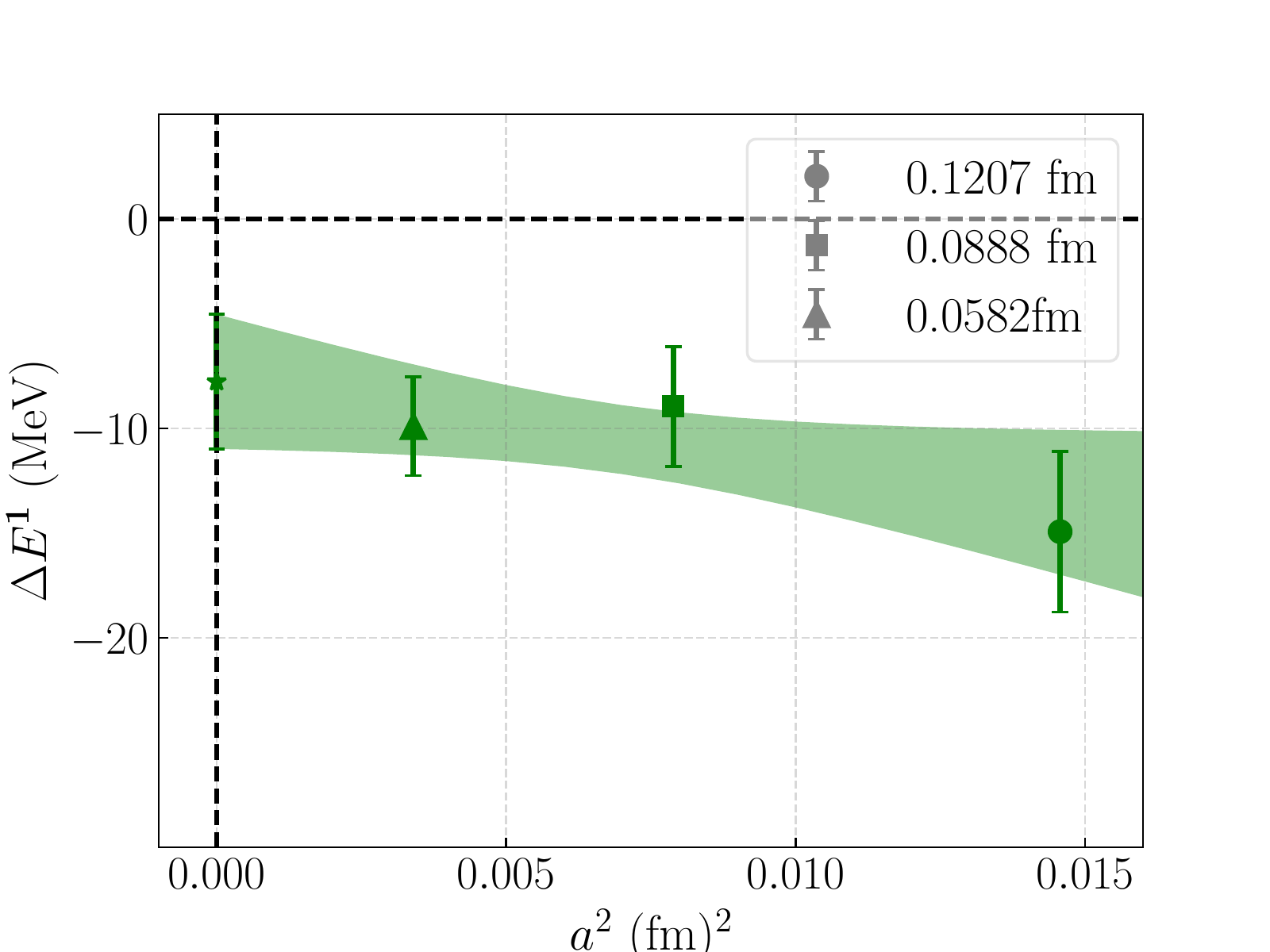}
	\caption{\label{fig:scbb} Continuum extrapolation of $sc\bar{b}\bar{b}$ state.}
\end{figure}
We begin with presenting data for the spin one doubly bottom tetraquark states.
As described earlier, we compute a matrix of correlation functions of the tetraquark $\mathcal{T}^{\mathbf{1}}(x)$ and two-mesons operators $\mathcal{M}^{\mathbf{1}}(x)$.
The diagonal correlators of this matrix correspond to the same source-sink operators while the off-diagonal correlators have a {\it tetraquark operator} at the source and a {\it two-meson operator} at the sink and vice-a-versa. 
The correlator matrix is non-hermitian and as mentioned earlier, in obtaining our final results, we employ  the eigenvector method of diagonalization.

As a representative plot on analysis, in Figure~\ref{fig:effmass} we show the effective mass of the lowest energy level obtained from such a diagonalization along with the  effective mass of the non-interacting two meson threshold correlator for the case of $us\bar{b}\bar{b}$.
The data in orange is the effective mass of the non-interacting two-meson correlator which in this case is obtained from the product of the correlators of the $B$ and $B^*_s$ mesons\footnote{\label{ft:1}In the case of the $us\bar{b}\bar{b}$ state, there exist two relevant threshold states namely $B B_s^*$ and $B_s B^*$. Of these two, we choose $B B_s^*$ which has relatively lower energy than that of $B_s B^*$. Similarly for all other flavor combinations, such as $uc\bar{b}\bar{b}$, $us\bar{c}\bar{c}$ and $sc\bar{b}\bar{b}$, we again choose the lowest strong decay threshold.}. The data in green is the effective mass of the lowest eigenvalue (the ground state) which is clearly below the effective mass of the threshold correlator. We also find that the effective mass corresponding to second eigenvalue overlaps with the effective mass of the threshold correlator in its approach to the plateau. However, as expected it is more noiser and need bigger basis of operators to extract it reliably.
As discussed previously, for each flavor combination we calculate the energy splitting $\Delta E^{\mathbf{1}}$ directly from Eq.~(\ref{Eq:eff_split}) by fitting the individual correlators as well as from the ratio of correlators using Eq.~(\ref{Eq:approach_4}).

Following the above procedure we calculate the energy splittings ($\Delta E^{\mathbf{1}}$) for all the doubly bottom tetraquarks with various flavor combinations mentioned in Table II.
This is performed on three different lattices ($a \sim 0.12, 0.09$ and $0.06$ fm) and on each one we vary the light quark masses over a wide range as listed in Table~\ref{tab:spinone}.
In Figure~\ref{fig:uqbb_1}, we show these results where in the left panel we plot these energies computed at various pion masses.
The results for the flavor combinations, $uq\bar{b}\bar{b}$ with $q \in (d,s,c)$ are shown by red, green and blue colored data, respectively.
As a representative plot we choose to show results at the coarse lattice spacing since here we have the maximum number of pion masses and therefore can show the pion mass dependence of these energy splittings ($\Delta E^{\mathbf{1}}$) more prominently. Result for the $ud\bar{b}\bar{b}$ state exhibits larger uncertainties at lower pion masses due to the presence of two light  quarks while the state $us\bar{b}\bar{b}$ allows us to extract results at much lower pion masses. For $uc\bar{b}\bar{b}$ we could extract results even at the physical light quark mass. 

It can be noted that for all the flavor combinations, there is a trend of increment of $\Delta E^{\mathbf{1}}$ with the lowering of pion masses and we will discuss the details shortly.
The availability of a large number of data points allows us to perform the chiral extrapolation much reliably. 
At each lattice spacing, we first perform the chiral extrapolation of $\Delta E^{\mathbf{1}}$ and then perform a continuum extrapolation from the results obtained at three lattice spacings. We use the following simple quadratic ansatz for both chiral and continuum extrapolations: 
\begin{eqnarray}
\Delta E^k_{m_\pi} &= c^{k}_1 + c^{k}_2\  m^2_\pi,\label{Eq:chiral} \\
\Delta E^k_{a} &= c^{k,a}_1 + c^{k,a}_2\  a^2. \label{Eq:conti} 
\end{eqnarray}
Here the label $k$ for the spin is kept general since we will also use these ansatz for both spin sectors. We perform two fittings: one including all data points to show the pion mass dependence over a wide range of pion masses and the other with only the lower few pion masses to perform the chiral extrapolation. The fit results are shown in Table~\ref{tab:chiral}, where in the second column we show the relevant slope parameter labelled as $c^{\mathbf{1},m_\pi}_2$ which is indicative of the pion mass dependence of the energy splitting $\Delta E^{\mathbf{1}}$. 
It is instructive to compare $c^{\mathbf{1},m_\pi}_2$ parameters for different tetraquark states with different flavor combinations at a given lattice spacing. The fits indicate that the state $ud\bar{b}\bar{b}$ exhibits the most pronounced trend in the increase of $\Delta E^{\mathbf{1}}$, followed by the state $us\bar{b}\bar{b}$ while the state $uc\bar{b}\bar{b}$ exhibits a very minute variation.
The results at the finest lattice spacings do not indicate such a clear trend as we do not have data points at much lighter pion masses at this lattice spacing.
{
\begingroup
\renewcommand*{\arraystretch}{1.6}
\setlength{\tabcolsep}{4pt}
\begin{table}[ht]
	\centering
	\caption{\label{tab:chiral}{Pion mass dependence and chiral extrapolation results for the spin one tetraquarks with different flavor combinations on three different lattices.}}
	\begin{tabular}{c c c c c c }\hline \hline
		State & $a$  &$c^{\mathbf{1},\pi}_2$ & $m^{\text{cut}}_\pi$ & $c^{\mathbf{1},\text{chiral}}_2$  & $\Delta E^{\mathbf{1}} |^{m_\pi^{\text{phys}}}$ \\
		& (fm) & (MeV) & & & (MeV) \\ \hline \hline
		\multirow{3}{*}{$ud\bar{b}\bar{b}$} & 0.1207  & 165(40) & 539 & 152(76)   & -158.1(18.0) \\
						    & 0.0888  & 246(71) & 688 & 246(71)   & -171.9(27.4)  \\
						    & 0.0582  & 102(56) & 645 & 102(85)  & -134.3(29.6) \\
		\hline
		\multirow{3}{*}{$us\bar{b}\bar{b}$} & 0.1207  & 80(13) & 297  & 82(376)   & -121.2(16.4) \\
						    & 0.0888  & 91(55) & 537  & 130(133)  & -108.8(28.5)  \\
						    & 0.0582  & 21(53) & 645  & 3(80)       & -93.1(27.8) \\
		\hline
		\multirow{3}{*}{$uc\bar{b}\bar{b}$} & 0.1207  & 30(9) & 257  & 183(306)  & -33.3(10.9) \\
						    & 0.0888  & 21(14)& 441  & 71(89)      & -24.6(12.1)  \\
	                                            & 0.0582  & 6(17) & 645  & 3(25)       & -12.0(8.6) \\
		\hline \hline
		\multirow{3}{*}{$ud\bar{c}\bar{c}$} & 0.1207  & 54(10) & 449  &  44(28) & -31.4(5.8) \\
					            & 0.0888  & 43(17) & 688  &  43(17)  & -31.9(6.6)  \\
	                                            & 0.0582  & 8(18)  & 688  &  9(34)   & -18.5(11.9) \\
		\hline
		\multirow{3}{*}{$us\bar{c}\bar{c}$} & 0.1207  & 4(6)   & 449  &  -8(9)  & -11.4(2.5) \\
	                                            & 0.0888  & -7(11) & 537  & -31(30) & -10.2(3.8)  \\
	                                            & 0.0582  & -7(17) & 688  &  -7(17) & -11.0(6.6) \\
		\hline \hline
	\end{tabular}
\end{table}
\endgroup
\begingroup
\renewcommand*{\arraystretch}{1.9}
\setlength{\tabcolsep}{5pt}
\begin{table}[ht]
	\centering
	\caption{\label{tab:conti}{Continuum extrapolation results for the various flavors of tetraquark states in the spin one sector. The fourth column is the continuum extrapolation results from three lattices. The last column is obtained by averaging results from coarser two lattices.}}
	\begin{tabular}{c c c c | c }\hline \hline
	State & $c^{\mathbf{1},a}_1$ & $c^{\mathbf{1},a}_2$  & $\Delta E^{\mathbf{1}}|_{a=0}^{m_\pi^{\text{phys}}}$ & $\Delta E^{\mathbf{1}}|_{\text{avg}}$\\ 
		 &       &      &   (MeV)  & (MeV) \\    \hline \hline
	$ud\bar{b}\bar{b}$ & -143(34) & -1239(2915)  & -143.3(33.9) & -165.0(32.5)\\
	$us\bar{b}\bar{b}$ & -87(32)  & -2393(2725)  & -86.7(32.4) & -115.0(32.8) \\
	$uc\bar{b}\bar{b}$ & -6(11)   & -1918(1239)  & -6.4(11.2) & -28.95(16.3) \\
	$sc\bar{b}\bar{b}$ & -8(3)    & -395(398)    & -7.67(3.21) & -11.94(4.7)\\ 
	\hline
	$ud\bar{c}\bar{c}$ & -23(11) & -637(1001)  & -23.3(11.4) & -31.7(8.8) \\
	$us\bar{c}\bar{c}$ & -8(8) & -241(574)    & -7.7(7.5) &  -10.8(4.5) \\
	\hline \hline	
	\end{tabular}
\end{table}
\endgroup 
} 

For the second fit, {\it i.e.}, for the chiral extrapolation, we use the ansatz in Eq.~(\ref{Eq:chiral}) and employ cuts on the largest pion masses and include data  corresponding to as low pion masses as can be afforded by meaningful uncertainties in the extrapolation.
The results of the chiral extrapolation are shown in Table~\ref{tab:chiral} with the appropriate slope parameter labelled as $c^{\mathbf{1},\text{chiral}}_2$ in column 5, and the relevant maximum pion mass used in the fit being labelled as $m^{\text{cut}}_\pi$ is shown in column 4.
The chirally extrapolated values of  $\Delta E^{\mathbf{1}} |_{m^{\text{phys}}_\pi}$ are shown in the last column.
We then use these chirally extrapolated $\Delta E^{\mathbf{1}} |_{m^{\text{phys}}_\pi}$ from three different lattice spacings and perform a continuum extrapolation using the ansatz in Eq.~(\ref{Eq:conti}).
The results of this extrapolation are shown in the right panel of Figure~\ref{fig:uqbb_1} and the fit results are listed in Table~\ref{tab:conti}.
The slope parameter $c^{\mathbf{1},a}_2$ in this case will be an indicator of the lattice spacing dependence of the particular state.
For $ud\bar{b}\bar{b}$ and $us\bar{b}\bar{b}$, these are  consistent with zero indicating no dependence on lattice spacing.
The parameter $c^{\mathbf{1},a}_2$ for $uc\bar{b}\bar{b}$ state indicates a mild dependence on the lattice spacing. The state $sc\bar{b}\bar{b}$, which is the SU(3) symmetric state of $uc\bar{b}\bar{b}$, requires no chiral extrapolation since all quark masses are at their physical values. The corresponding lattice spacing dependence parameter, $c^{\mathbf{1},a}_2$,  as shown in Table~\ref{tab:conti}, indicates no dependence on lattice spacing of this state. The continuum extrapolated results $\Delta E^{\mathbf{1}}|_{a=0}^{m_\pi^{\text{phys}}}$ are shown in Figure~\ref{fig:scbb}.

It can be noted that at the finest lattice spacing, the lowest pion mass available is $m_\pi = 545$ MeV, which may not be low enough for a chiral extrapolation.
Because of this reason, the chirally extrapolated results at this lattice spacing may have a systematic effect arising from the absence of lower pion masses and that may reflect in the lattice spacing dependence of some of our findings such as for $uc\bar{b}\bar{b}$ state. 
Hence we also report our results without including data from the fine lattice. 
Since we are left with only two data points, we have not performed any fit (with 2 degrees of freedom) in this case. 
Instead we average the results obtained on other two lattices (with spacings 0.0888 and 0.1207 fm) and report that with errorbars added in the quadrature. 
In column 5 of Table~\ref{tab:conti} we show those average results by $\Delta E^{\mathbf{1}}|_{\text{avg}}$.

We now discuss the results of the spin one doubly charm tetraquarks.
In Figure~\ref{fig:uqcc} we show those results where the left panel shows the pion mass dependence and the chiral extrapolation on the coarse lattice. 
The right panel represents results for the continuum extrapolation.  
The relevant lowest thresholds for the flavor combinations $ud\bar{c}\bar{c}$ and $us\bar{c}\bar{c}$  are the non-interacting $D$-$D^*$ and $D$-$D^*_s$ mesons, respectively.
For both cases, we find an energy level below their relevant strong decay thresholds while the other energy level appears at the threshold. 
As in the doubly bottom cases, we calculate the energy splittings ($\Delta E^{\mathbf{1}}$ in Eq.(\ref{Eq:eff_split})) between the lowest energy levels and the threshold states by direct fitting as well as from the ratio of correlators (as in Eq.(\ref{Eq:approach_4})). 
We represent the fitted results for $ud\bar{c}\bar{c}$ by red data points while results for $us\bar{c}\bar{c}$ are shown by green points. 
The fitted results for pion mass dependence and chiral extrapolation are shown  in Table~\ref{tab:chiral}, while the results for continuum extrapolation are shown in  Table~\ref{tab:conti}.
In the case of $ud\bar{c}\bar{c}$, similar to $ud\bar{b}\bar{b}$, we observe a trend in the increase of $\Delta E^{\mathbf{1}}$ with the lowering of the light quark constituents. This is evident from the fits for the pion mass dependence and is indicated by $c^{\mathbf{1},m_\pi}$ parameter
on the coarsest two lattice spacings.
The finest lattice spacing results do not clearly indicate this trend due to the lack of lower pion masses at that lattice spacing. 
The pion mass dependence of the energy splitting for $us\bar{c}\bar{c}$, color coded in green, is much flatter in comparison to $ud\bar{c}\bar{c}$ and this trend is reflected in the $c^{\mathbf{1},m_\pi}_2$ coefficient. The continuum extrapolations for both $ud\bar{c}\bar{c}$ and $us\bar{c}\bar{c}$ indicate no discernible dependence on the lattice spacing. 

In column 4 of Table~\ref{tab:conti},  we show the continuum extrapolated results for doubly charmed tetraquarks. The column 5 shows the average results obtained on coarse two lattices. 
Both columns show the presence of energy levels below their respective thresholds both for $ud\bar{c}\bar{c}$ and $us\bar{c}\bar{c}$. However, they are very close to their respective strong decay thresholds as was also observed in Ref. \cite{Cheung:2017tnt}.
Because of their close proximity to thresholds, a careful finite volume analysis \cite {Luscher:1990ck} is needed to make conclusive statements about the nature of these states.
Though they could be stable under strong interaction they may not appear as bound states because of threshold effects.

\begin{figure}
	\includegraphics[width=1.0\linewidth]{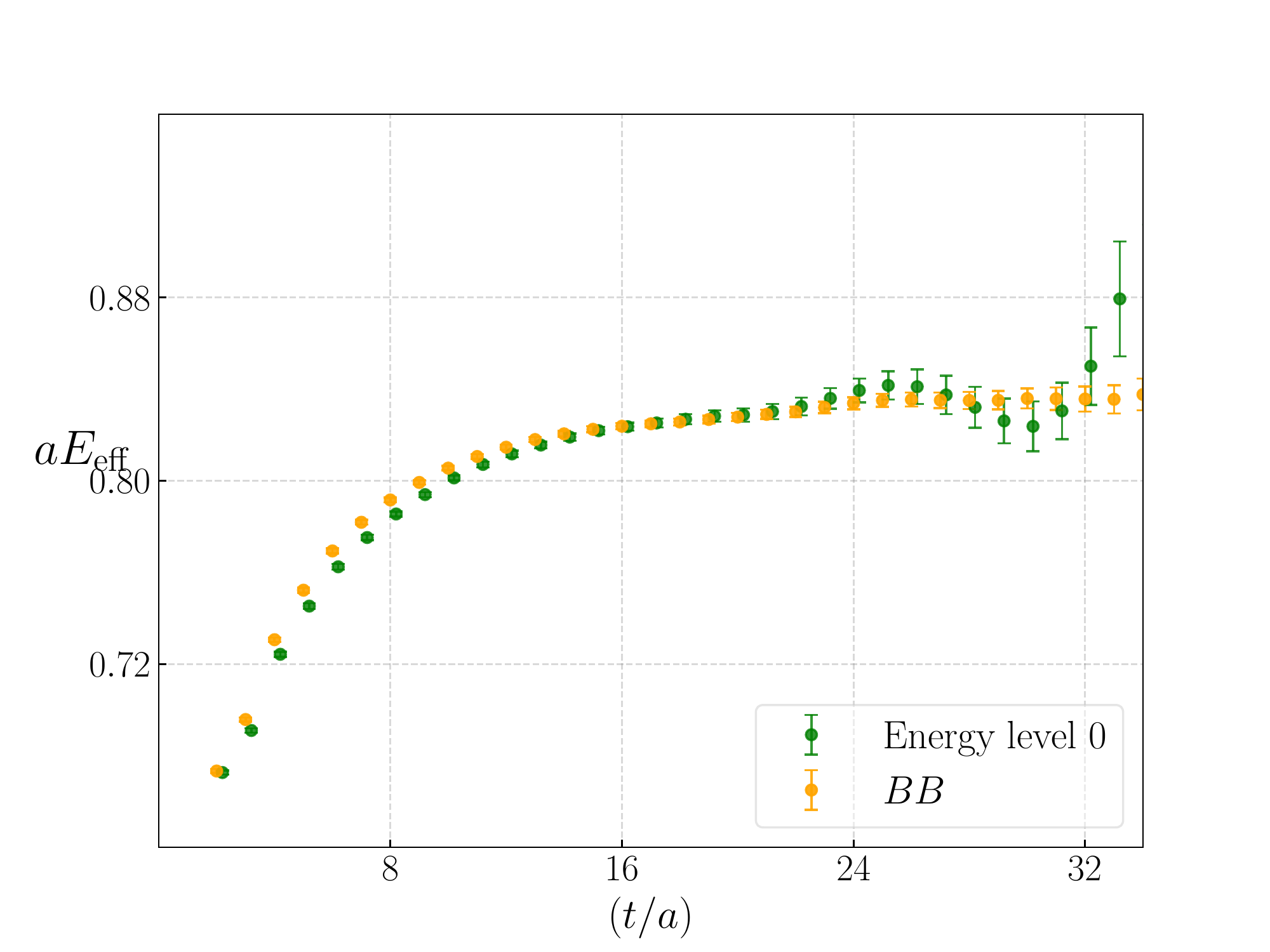}
	\caption{\label{fig:effmass_uubb} Effective mass of the ground state energy level (green) obtained from the the GEVP solution for the spin 0, $uu\bar{b}\bar{b}$ tetraquark state. The data in orange is the effective mass of the threshold correlator $BB$. Results computed at $a=0.0583$ fm and at $m_\pi= 688$ MeV.}
\end{figure}
\begin{figure*}[ht]
	\includegraphics[width=1.0\linewidth]{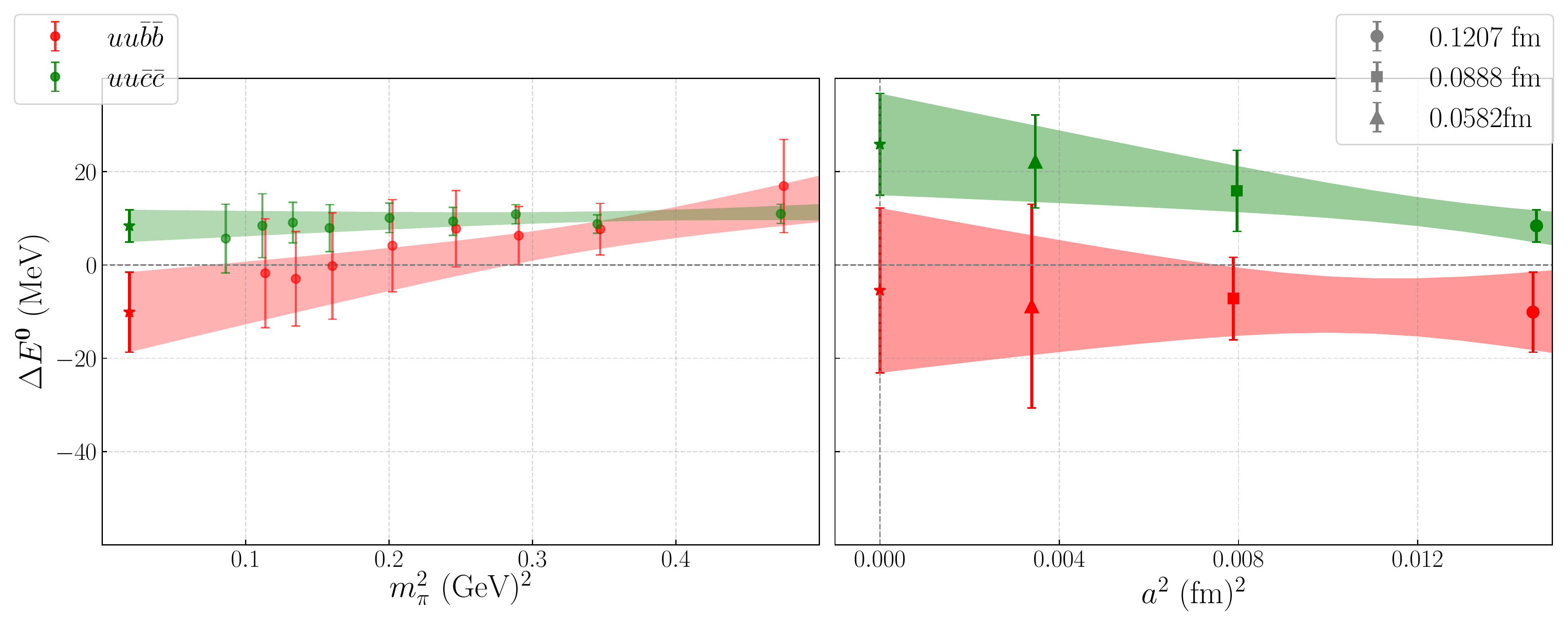}
	\caption{\label{fig:llbb_1} Results of spin zero $uu\bar{b}\bar{b}$ and $uu\bar{c}\bar{c}$ tetraquark states.
	 Left panel: Energy splittings at several pion masses at $a = 0.1207$ fm for both the states. The fit bands indicate a chiral extrapolation fit as per Eq.~(\ref{Eq:chiral}) color coded appropriately for each state. Right panel: Continuum extrapolation results as per Eq.~(\ref{Eq:conti}) from three lattice spacings. The data point at each lattice spacing is the result of the chiral extrapolation to the physical pion mass at that lattice spacing.}
\end{figure*}
\begin{figure*}[ht]
	\includegraphics[width=1.0\linewidth]{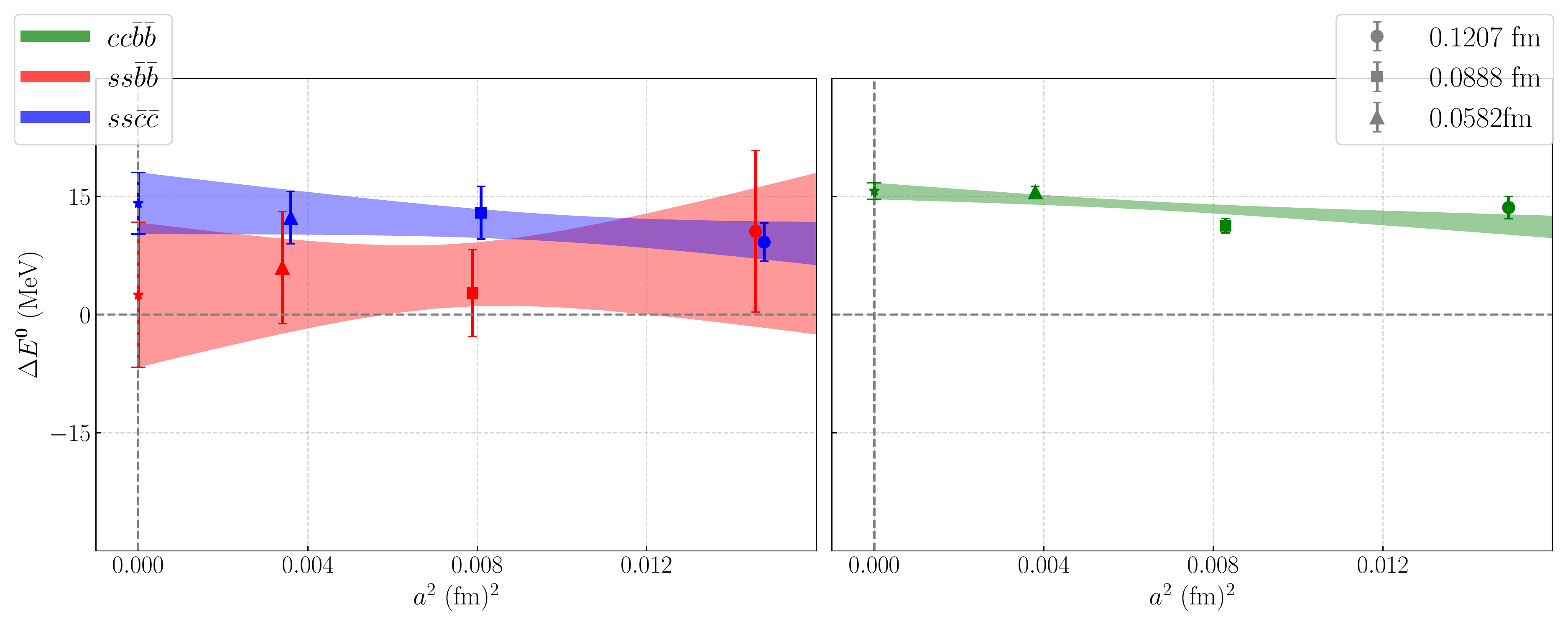}
	\caption{\label{fig:llbb_2} Left:  Continuum extrapolation of $ss\bar{b}\bar{b}$ and $ss\bar{c}\bar{c}$ states from three lattice spacings. Right: Continuum extrapolation of the $cc\bar{b}\bar{b}$.}
\end{figure*}

\subsection{Spin zero tetraquarks $J^P=0^+$\label{subsec:spinzero}}
In the spin zero sector, we compute the energy levels of the tetraquark states with various flavor combinations that are listed in Table~\ref{tab:spinzero}.
These tetraquark states are flavor symmetric cousins of those listed in Table~\ref{tab:spinone}.
As in the case of spin one sector, we compute a matrix of correlation functions  consisting {\it tetraquark-type}, $\mathcal{T}^{\mathbf{0}}(x)$, and {\it two-meson-type}, $\mathcal{M}^{\mathbf{0}}(x)$, interpolating fields and employ the eigenvector method of diagonalization in obtaining our final results. 

We shall begin by discussing the spin zero doubly charmed and doubly bottom tetraquark states with $I=1$.
The effective masses of the principal correlators, obtained from GEVP analysis, for the flavor combination $uu\bar{b}\bar{b}$ are shown in  Figure~\ref{fig:effmass_uubb}. This representative figure is obtained on the fine lattice and at the pion mass $m_\pi= 688$ MeV. 
The relevant strong decay threshold in this case is the two non-interacting $B$ mesons.
The effective mass of the product correlator of two $B$ mesons is represented by the orange data. 
The effective mass of the lowest eigenvalue, shown in green, is seen to coincide with the threshold correlator.
This behavior is in contrast when compared with its flavor anti-symmetric partner $ud\bar{b}\bar{b}$ where there is a clear indication of the ground state level being below the relevant threshold.
The energy splitting ($\Delta E^{\mathbf{0}}$ in Eq.~(\ref{Eq:eff_split})) of the tetraquark state $uu\bar{b}\bar{b}$, is shown at the left panel of Figure~\ref{fig:llbb_1}  by red colored data points where results are obtained at various pion masses (on the coarser lattice) to explore the pion mass dependence.
We note that the determination of these energy splittings is significantly noisier in comparison to the spin one $ud\bar{b}\bar{b}$ state with the same statistics. 
This limits us in using much lighter pion masses for $uu\bar{b}\bar{b}$. 
Furthermore, this also forces us to use the entire dataset for exploring both the pion mass dependence as well as the chiral extrapolation.
We perform a chiral extrapolation with the ansatz in Eq.~(\ref{Eq:chiral}) at each lattice spacing and the results are listed in Table~\ref{tab:chiral_llbb}.
The fits for the parameter $c^{\mathbf{0},m_\pi}_2$ indicate a dependence on pion mass for $a=0.1207$ fm and no dependence is seen for the other two lattice spacings, since $c^{\mathbf{0},m_\pi}_2$ is consistent with zero.
It can be noted that this behavior again is in contrast with the pion mass dependence of the $ud\bar{b}\bar{b}$ state where a non-trivial dependence was clearly identified.
After the chiral extrapolation, we perform the continuum extrapolation using the ansatz in Eq.~(\ref{Eq:conti}) and fits are shown in Table~\ref{tab:conti_llbb}.
The slope parameter $c^{\mathbf{0},a}_2$ for the state $uu\bar{b}\bar{b}$ is consistent with zero indicating no dependence on the lattice spacing.
The physical and continuum extrapolated result for $uu\bar{b}\bar{b}$ clearly indicates that there is no energy level below its lowest strong decay threshold with any statistical significance and is consistent with zero.

The green data points in Figure~\ref{fig:llbb_1} show the results for $\Delta E^{\mathbf{0}}$ (on  $a=0.1207$ fm lattice) for the spin zero doubly charmed tetraquarks $uu\bar{c}\bar{c}$.
In this case the GEVP solutions also display similar qualitative features as the corresponding doubly bottom states where the ground state coincides with the threshold and a well separated second state lies above that.
Here, the threshold is that of the two non-interacting $D$ mesons.
As in the previous case, we use the entire dataset for the pion mass dependence as well as chiral extrapolation.
The chiral extrapolation fits at each lattice spacing shown in Table~\ref{tab:chiral_llbb} indicate no dependence on the pion mass since the parameter $c^{\mathbf{0},m_\pi}_2$ is found to be consistent with zero.
The continuum extrapolation for this case, color coded in green, is shown in the right panel of Figure~\ref{fig:llbb_1}, which indicates a mild dependence on the lattice spacing.
The physical and continuum extrapolated results ($\Delta E^{\mathbf{1}}|_{a=0}^{m_\pi^{\text{phys}}}$) are shown in the fifth column of Table~\ref{tab:conti_llbb} and all are found to lie above the respective threshold states.  
As in the spin one case, we have also calculated the average values of these energy splittings from the results obtained on two coarse lattices, and show that in the last column of Table~\ref{tab:conti_llbb}.

With our available quark propagators we are also able to study $I= 0, J = 0$ tetraquark states, $ss\bar{b}\bar{b}, ss\bar{c}\bar{c}$ and $cc\bar{b}\bar{b}$, where the strange, charm and bottom quark masses are tuned to their physical values. 
Energy levels obtained for these states will thus be at the physical points and there is no need for any chiral extrapolation. 
The thresholds for these states are the non-interacting $B_s B_s$, $D_s D_s$ and $B_c B_c$, respectively.
These require only a continuum extrapolation which are shown in the two panels of Figure~\ref{fig:llbb_2}, and the fitted results are shown in Table~\ref{tab:conti_llbb}.
The estimates of the energy splitting $\Delta E^{\mathbf{0}}$ for the state $ss\bar{b}\bar{b}$ (color coded in red) show no lattice spacing dependence and the final result is consistent with zero indicating the absence of any bound state.
For the state $ss\bar{c}\bar{c}$ we also find similar results and the continuum extrapolated result lie above its respective threshold which is most likely to be a scattering state. Results for the state $cc\bar{b}\bar{b}$ indicates a mild lattice spacing dependence and the continuum result is also most likely be a scattering state. In conclusion, our analysis on the $I=0$, spin zero, tetraquarks with flavor combinations $ss\bar{b}\bar{b}, ss\bar{c}\bar{c}$ and $cc\bar{b}\bar{b}$ suggest the absence of any bound state and the observed energy levels correspond to the scattering states.
Recently a potential based lattice QCD study in Ref.~\cite{Bicudo:2015vta} for doubly bottom spin zero states also concluded the same.
{
\begingroup
\noindent
\renewcommand*{\arraystretch}{1.6}
\setlength{\tabcolsep}{4pt}
\begin{table}[t]
	\centering
	\caption{\label{tab:chiral_llbb}{Chiral continuum extrapolation results for various lattice spacings and flavors of tetraquark states in the spin zero sector.}}
	\begin{tabular}{c c c c c }\hline \hline
		State & $a$ &$c^{\mathbf{0},m_\pi}_1$ & $c^{\mathbf{0},m_\pi}_2$  & $\Delta E^{\mathbf{0}} |^{m_\pi^{\text{phys}}}$ \\
		& (fm) & & & (MeV)\\ \hline \hline
		\multirow{3}{*}{$uu\bar{b}\bar{b}$} & 0.1207 & -11(9) & 50(25) & -10.1(8.6) \\
																& 0.0888  & -8(9)  & 26(26) & -7.2(8.8)  \\
																& 0.0582  & -9(23) & 33(57) & -8.8(21.9) \\
		\hline
			\multirow{3}{*}{$uu\bar{c}\bar{c}$} & 0.1207 &  8(4) & 6(10)   & 8.4(3.4) \\
																	& 0.0888  & 16(9) & -10(22) & 15.9(8.7) \\
																	& 0.0582 & 22(10) & -12(24)& 22.2(10)  \\
												
		\hline \hline
	\end{tabular}
\end{table}
\endgroup
}
\begingroup
\noindent
\renewcommand*{\arraystretch}{1.9}
\setlength{\tabcolsep}{5pt}
\begin{table}[t]
	\centering
	\caption{\label{tab:conti_llbb}{Continuum extrapolation results for the various flavors of tetraquark states in the spin zero sector. The fourth
column is the continuum extrapolation results from three lattices.
The last column is obtained by averaging results from
coarser two lattices.}}
	\begin{tabular}{c c c c | c}\hline \hline
	State & $c^{\mathbf{0},a}_1$ & $c^{\mathbf{0},a}_2$  & $\Delta E^{\mathbf{0}}|_{a=0}^{m_\pi^{\text{phys}}}$ & $\Delta E^{\mathbf{0}}|_{\text{avg}}$\\
	      &         &          &     (MeV) & (MeV) \\ \hline \hline
	$uu\bar{b}\bar{b}$ & -5(18) & -303(1549) & -5.5(17.7) & -8.7(12.3) \\
	$uu\bar{c}\bar{c}$ &  26(11) & -1202(824)  & 25.9(10.9) & 12.15(9.3)\\ 
	\hline
	$ss\bar{b}\bar{b}$ & 3(9) & 328(1108) & 2.5(9.2) & 6.6(11)\\
	$ss\bar{c}\bar{c}$ &  14(4) & -319(356)  & 14.1(3.9) & 11.1(4.1)\\ 
	$cc\bar{b}\bar{b}$ &  16(1) & -285(139)  & 15.7(1.0) & 12.5(1.69) \\
	\hline \hline	
	\end{tabular}
\end{table}
\endgroup

\subsection{Finite volume effects}
For all the spin one tetraquark states with various flavor combinations listed in Table~\ref{tab:spinone}, we have found the energy levels below their respective strong decay thresholds. 
In some cases the energy splittings ($\Delta E^{\mathbf{1}}$) between the ground state and the threshold state are very large while for others they are close and below their respective thresholds. 
However, all these energy levels are obtained within a single volume of about 3 fm. 
It is thus necessary to estimate the finite volume effects on these energy differences and obtain their infinite volume estimates which can then be interpreted as the binding energies of the corresponding bound states.
However, repeating these calculations on multiple lattice volumes is computationally very expensive and so is beyond the scope of this work. 

However, it is possible to identify a few states for which the finite volume corrections will be suppressed, {\it i.e.}, could be very small. The estimation of $\Delta E^{\mathbf{1}}$ on single large enough volume for such a case, in fact, would be close to its binding energy ($B_{\infty}$). 
As demonstrated  in references \cite{Beane:2003da,Davoudi:2011md,Briceno:2013bda}, the finite volume corrections $\Delta_{FV}$ to energy levels corresponding to an infinite volume bound state with energy $E_{\infty}$ scale as, 
\begin{eqnarray}\label{Eq:finite_vol_eff}
 \Delta_{FV} =  E_{FV} - E_{\infty} &\propto &  \mathcal{O}(e^{-k_{\infty} L })/L,\nonumber \\
   \mathrm{with}\quad k_{\infty} &=& \sqrt{(m_1 + m_2) B_{\infty}}\,,
\end{eqnarray} 
where, $E_{FV}$ is the energy level computed a cubic lattice, $k_{\infty}$ is the binding momentum of the infinite volume state and ($m_1, m_2$) are the masses of the two non-interacting particles with the threshold energy $m_1+m_2$.  
It should be noted from the above expression that the finite volume effects are suppressed by the threshold mass ($m_1+m_2$) and that this suppression is significantly enhanced for the cases where the threshold states are heavy mesons, such as those we are studying here. 
In addition to that if $\Delta E$ is also large, then the finite volume corrections will further be suppressed since it also enters in the exponential. 
Therefore in the doubly bottom sector, tetraquark states with the flavor combinations, $ud\bar{b}\bar{b}$ and $us\bar{b}\bar{b}$, for which the $\Delta E$ values are found to be
more than 150 and 100 MeV, respectively, will have small finite
volume corrections. 
For these cases it is quite natural to expect that the energy splitting $\Delta E$ will be closer to their infinite volume binding energy.
Therefore these states will be stable under strong interactions. 
However, for the cases, particularly for the doubly charmed tetraquarks, which are below but closer to their thresholds ({\it i.e.}, $\Delta E$ values are closer to zero), it will be difficult to get any qualitative estimate for their finite volume corrections. In those cases one needs to perform a detail finite volume
study \cite{Luscher:1990ck} to make any conclusive statement about
their infinite volume pole structures.

\begin{figure*}[ht]
	\includegraphics[width=1.0\linewidth]{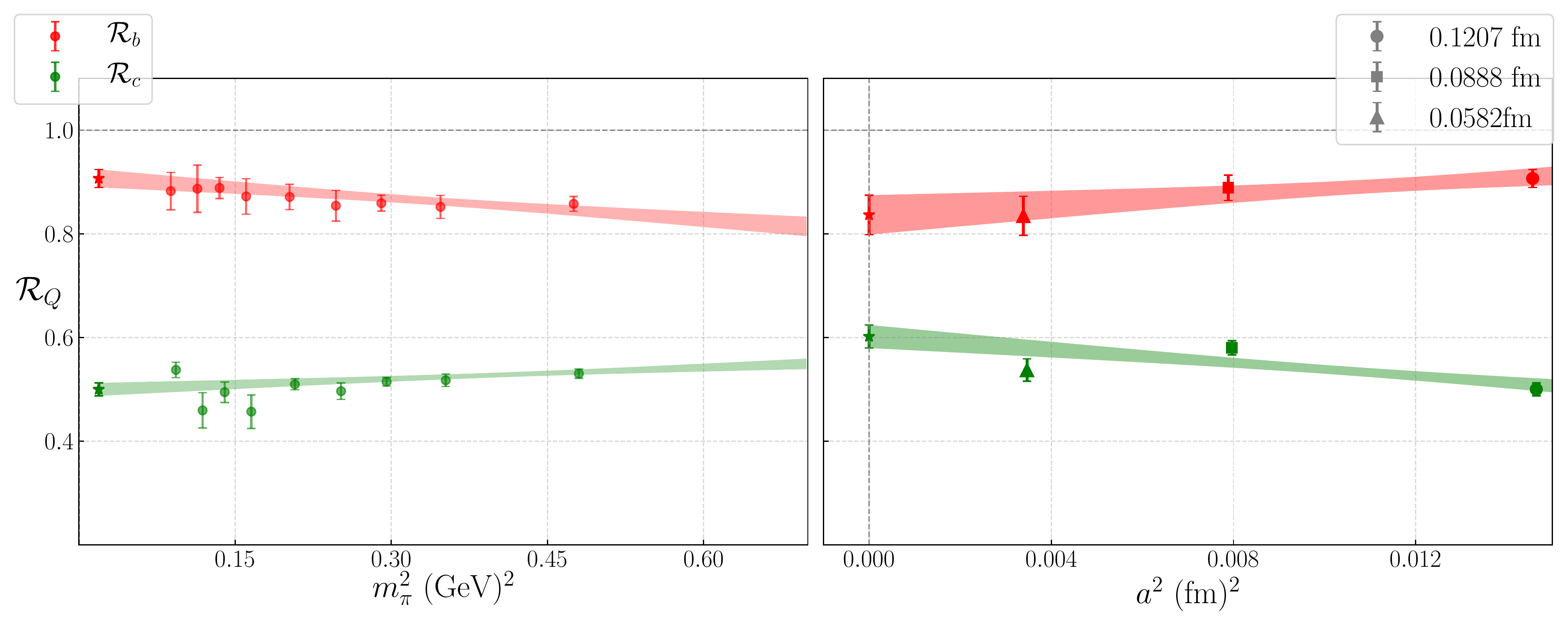}
	\caption{\label{fig:hqet_test} Results of the ratios $\mathcal{R}_{b}$ and $\mathcal{R}_c$ color coded as red and green respectively. Left:  Results of the chiral extrapolation at $a=0.1207$ fm. Right: Continuum extrapolation results from three lattice spacings.}
\end{figure*}

\subsection{Heavy quark effective theory and hadron mass relations}\label{subsec:hqet_test}
The Heavy quark effective theory (HQET) is a very useful tool and is often utilized to understand various properties of heavy hadrons including their energy spectra. Using heavy quark symmetries one can also obtain mass relations between heavy flavored hadrons such as those mentioned in Ref.~\cite{Eichten:2017ffp}.
Using such symmetry relations, Ref.~\cite{Eichten:2017ffp} predicted masses and binding energies of various tetraquarks states including some of those studied in this work. 
Although such relations are valid in the infinite quark mass limit, they are used at the bottom and even at the charm quark masses. 
It will therefore be interesting to investigate these relations by a first principles non-perturbative method, such as lattice QCD, with a goal to validate these relations at a given quark mass and access their deviation, if any, from the heavy quark limit.
The availability of data on the ground state masses on mesons, baryons and tetraquarks obtained from this calculation, both at the charm and the bottom quark masses, provides such an opportunity to systematically investigate these relations. Below we elaborate that.


The work in Ref.~\cite{Eichten:2017ffp} states the following relation amongst the hadrons with heavy quarks:
\begin{eqnarray}\label{Eq:hqet_relation}
  m(\{ Q_i Q_j\}[\bar{q}_k \bar{q}_l]) &-& m(\{ Q_i Q_j\} q_y)\nonumber \\
    &=& m(Q_x[q_k q_l]) - m(Q_x \bar{q}_y),
\end{eqnarray} 
where $Q_i, q_k$ denote heavy and light quarks respectively. Here we use the same notation as in Ref. \cite{Eichten:2017ffp}. The braces $\{ ... \}$ and $[...]$ imply the symmetrization and anti-symmetrization, respectively, with respect to the flavor degrees of freedom.
In this notation, $(\{ Q_i Q_j\}[\bar{q}_k \bar{q}_l])$\footnote{The tetraquark operator used in this work is a complex conjugate of this operator.} represents a tetraquark operator with the  flavor symmetries indicated by the braces, while $(\{ Q_i Q_j\} q_y)$, $(Q_x[q_k q_l])$ and $(Q_x \bar{q}_y)$ represent a heavy-heavy-light baryon, heavy-light-light baryon and heavy-light meson respectively.
It should be noted that Ref.~\cite{Eichten:2017ffp} provides four such relations depending on the combination of flavor symmetrization/anti-symmetrization and the one shown here corresponds to our operator construction.
The relation in Eq.~(\ref{Eq:hqet_relation}) can then be employed to predict the masses of the tetraquark states by substituting the relevant masses of heavy baryons and mesons.
In Ref.~\cite{Eichten:2017ffp} this was calculated by using the spin average masses of the charmonia, bottomonia and heavy baryons by inserting their experimental or quark model values. 

%

Here, we aim to study this relation both at the charm and the bottom quark masses. We do not consider the spin-average mass, instead use the spin-1/2 states for baryons and pseudoscalar mass for the heavy-light meson. If there is any deviation from the equality for Eq.~(\ref{Eq:hqet_relation}) that would be maximum in this choice.  
In doing so, we will be able to estimate an upper bound of the deviation from the heavy quark limit which originates from all $(1/m_Q)^n$ corrections. 
In evaluating Eq.~(\ref{Eq:hqet_relation}), we find it to be convenient{\footnote{The use of the ratio of masses allows for the cancellation of lattice artifacts in addition to the cancellation of uncertainties from resampling.}}  to redefine the relation as a ratio which for the charm and bottom quarks are given by:
\begin{equation}\label{Eq:hqet_ratio}
\mathcal{R}_b \equiv \frac{M_{ud\bar{b}\bar{b}} - M_{\Xi_{bb}}}{M_{\Lambda_b} - M_{B}}, \quad \mathcal{R}_c \equiv \frac{M_{ud\bar{c}\bar{c}} - M_{\Xi_{cc}}}{M_{\Lambda_c} - M_{D}}.
\end{equation}
In the  limit of infinitely heavy quarks, the ratio $\mathcal{R}_Q$ will be unity.
In computing these ratios ($\mathcal{R}_{c/b}$) we first evaluate the jackknife ratios of the following correlators:
\begin{eqnarray}\label{Eq:ratio_hqet}
\frac{C_{ud\bar{b}\bar{b}}(t)}{C_{\Xi_{bb}}(t)} &\to& A^\prime e^{-(M_{ud\bar{b}\bar{b}} - M_{\Xi_{bb}})t} + ... , \nonumber \\
\frac{C_{\Lambda_Q}(t)}{C_{M_{Q\bar{q}}}(t)} &\to& B^\prime e^{-(M_{\Lambda_Q} - M_{Q\bar{q}})t} + ... ,
\end{eqnarray}
which directly provide the difference of masses as shown above. $\mathcal{R}_{c/b}$ are then evaluated from the fits to these ratio correlators. 
In addition, we also fit the individual masses of tetraquarks, mesons and baryons and calculate $\mathcal{R}_{c/b}$ from Eq.~(\ref{Eq:hqet_ratio}). 
 We find consistent results with both methods and the evaluation with Eq.~(\ref{Eq:hqet_ratio}) provides improved uncertainties. 
 As we have access to a large number of light quark masses, while keeping the heavy quark mass at the charm and bottom quark, we vary the light quark mass and calculate $\mathcal{R}_{c/b}$ for each case.
In Figure~\ref{fig:hqet_test}, we show these results at several pion masses for the coarser lattice ($a \sim 0.12$ fm) using the entire dataset in fitting. This is done for other lattice spacings as well.
The results clearly indicate a wide separation of ratios between the charm and bottom quarks; while $\mathcal{R}_b$ is closer to the heavy quark limit of unity, $\mathcal{R}_c$ deviates from it substantially.
After repeating this calculation on other two lattices we perform a simplistic chiral and continuum extrapolation according to the ansatz in Eq.~(\ref{Eq:chiral}) and Eq.~(\ref{Eq:conti}). The fit results are shown in Tables~\ref{tab:chiral_hqet} and~\ref{tab:conti_hqet} at three lattice spacings.
For both ratios, $\mathcal{R}_b$ and $\mathcal{R}_c$, we do not observe any appreciable dependence on the pion mass as indicated by the parameter $c^\pi_2$ in Table~\ref{tab:chiral_hqet}. 
In addition, the continuum extrapolation fit in Table~\ref{tab:conti_hqet} do not indicate any lattice spacing dependence for the bottom and charm quarks.
The continuum extrapolated results are listed in the last column of Table~\ref{tab:chiral_hqet}; we find $\mathcal{R}_b = 0.837(38)$ and $\mathcal{R}_c = 0.602(22)$. 
These results clearly indicate that there is a substantial deviation from the heavy quark limit at the charm quark mass implying there might be a large contributions from  $(1/m_Q)^n$ corrections. 
However, results at the bottom quark mass are much closer to the heavy quark limit. Our results indicate that as far as the heavy quark symmetry relations such as that is shown in Eq.~(\ref{Eq:hqet_relation}) are considered, the charm quark mass is not heavy enough for the equality, and one certainly needs to incorporate appropriate leading order $1/m_Q$ and then higher order corrections terms. However, one can of course use these relations for bottom quarks with higher order $1/m_Q$ corrections.
{
\begingroup
\noindent
\renewcommand*{\arraystretch}{1.6}
\setlength{\tabcolsep}{4pt}
\begin{table}[t]
	\centering
	\caption{\label{tab:chiral_hqet}{Chiral extrapolation of ratios $\mathcal{R}_Q$} for charm and bottom quarks.}
	\begin{tabular}{c c c c c }\hline \hline
		Ratio & $a$ &$c^\pi_1$ & $c^\pi_2$  & $\mathcal{R}_Q |^{m_\pi^{\text{phys}}}$ \\
                \hline \hline
        \hline                       		
		\multirow{3}{*}{$\mathcal{R}_b$}    & 0.1207 & 0.91(2) & -0.14(5)  & 0.907(17) \\
														       & 0.088  & 0.89(3) & -0.03(0.1)& 0.889(24) \\
															   & 0.058  & 0.83(4) & 0.05(0.1) & 0.835(38) \\
		\hline
		\multirow{3}{*}{$\mathcal{R}_c$} & 0.1207 & 0.50(1) & 0.07(3)  & 0.500(13) \\
												    		& 0.088  & 0.58(1) & -0.05(5) & 0.580(14) \\
										 					& 0.058  & 0.54(2) & 0.03(6)  & 0.537(22)  \\
												
		\hline \hline
	\end{tabular}
\end{table}
\endgroup
}
\begingroup
\noindent
\renewcommand*{\arraystretch}{1.9}
\setlength{\tabcolsep}{5pt}
\begin{table}[t]
	\centering
	\caption{\label{tab:conti_hqet}{Continuum extrapolation of ratios $\mathcal{R}_Q$ for charm and bottom quarks.}}
	\begin{tabular}{c c c c }\hline \hline
	Ratio & $c^a_1$ & $c^a_2$  & $\mathcal{R}_Q|_{a=0}^{m_\pi^{\text{phys}}}$\\
        \hline \hline
	$\mathcal{R}_b$ & 0.84(4) & 5.01(3.18)  & 0.837(38) \\
	$\mathcal{R}_c$ & 0.60(2) & -6.33(2.01) & 0.602(22) \\ 
	\hline \hline	
	\end{tabular}
\end{table}
\endgroup 
\section{Discussion and Conclusions}~\label{sec:conclusions}
Recently there has been tremendous activities in studying multiquark states both theoretically and experimentally. In particular, heavy tetraquarks are being investigated at various laboratories as well as studied theoretically through different models and by lattice QCD calculations.
In this work, using lattice QCD 
we have performed a detailed study on the doubly heavy
tetraquark states with quark contents $q_1q_2\bar{Q}\bar{Q}, \, q_1,q_2 \subset u,d,s,c$ and $Q \equiv b,c$, in both spin zero ($J=0$) and spin one ($J=1$) sectors. Not only we study $ud\bar{b}\bar{b}$ and $us\bar{b}\bar{b}$, as was studied in Refs.~\cite{Francis:2016hui}, but also explore $uc\bar{b}\bar{b}, ud\bar{c}\bar{c}$ and $us\bar{c}\bar{c}$ states and additionally include the spin zero sector of doubly heavy tetraquarks. In doing so, we have presented a systematic dependence of the ground state spectra of such states on their light quark constituents over a wide range of quark masses starting from the quark mass corresponding to the physical pion mass to the strange quark mass. Since all these hadrons involve heavy quarks, naturally, like any heavy flavored hadrons, they are susceptible to heavy quark discretization effects in a lattice calculation. To check the lattice spacing dependence we have obtained results at three lattice spacings, finest one being at 0.0582 fm. At a given lattice spacing we perform a chiral extrapolation using several quark masses and then perform a continuum extrapolation to get the final results. For all the states in the spin one sector, we observe the presence of energy levels below their respective two-meson thresholds, deepest one being for the doubly bottom tetraquark, $ud\bar{b}\bar{b}$.
Furthermore, for various flavor combinations of the tetraquark states we find that there is a clear trend of increase in the energy splitting ($\Delta E$) as the light quark masses of such states are decreased and it becomes maximum at the physical quark mass. This energy splitting in the infinite volume limit of such a state can be interpreted as its binding energy. This trend was first indicated in the lattice calculation in Ref.~\cite{Francis:2016hui} for the states $ud\bar{b}\bar{b}$ and $us\bar{b}\bar{b}$. Here we confirm that over a wide range of quark masses. Additionally we find that such a trend holds for all the spin one states considered here including the doubly charm tetraquark states.
For the doubly charmed tetraquark states, $ud\bar{c}\bar{c}$ and $us\bar{c}\bar{c}$, we also find that the ground states are below their respective thresholds. However,
they are quite close to their thresholds which was also observed in Ref. \cite{Cheung:2017tnt}. Though
they could be stable under strong interactions one needs to carry out finite volume analysis to establish their bound state properties, if there is any. We would also like to point out that most of these states, except $uc\bar{b}\bar{b}$, show either no discernible dependence or very mild dependence on lattice spacing. However, this will be clear when in future study we include much lower pion masses on the fine lattice. Our final results for doubly heavy spin one tetraquarks states from this calculation are summarized in Table \ref{tab:spinone_final}. 
\begingroup
\renewcommand*{\arraystretch}{1.9}
\begin{table}[ht]
	\centering
	\caption{\label{tab:spinone_final}{Final results for the spin one tetraquarks}}
	\begin{tabular}{c c | c c  }\hline \hline 
		State &  $\Delta E^{\mathbf{1}}$ [MeV] & State &  $\Delta E^{\mathbf{1}}$ [MeV] \\ \hline \hline
		$ud\bar{b}\bar{b}$   & -143(34) & $us\bar{b}\bar{b}$  & -87(32) \\ \hline
		 $uc\bar{b}\bar{b}$  & -6(11)   & $sc\bar{b}\bar{b}$ & -8(3) \\ \hline
 		$ud\bar{c}\bar{c}$   & -23(11) & $us\bar{c}\bar{c}$   &  -8(8)\\ \hline \hline
	\end{tabular}
\end{table}
\endgroup
Our estimates for the $ud\bar{b}\bar{b}$ and $us\bar{b}\bar{b}$ are in agreement with those of Ref. \cite{Francis:2016hui} at a lattice spacing ($\sim$ 0.09 fm) where both of ours data are available.

\begin{figure}[h]
	\includegraphics[width=1.0\linewidth]{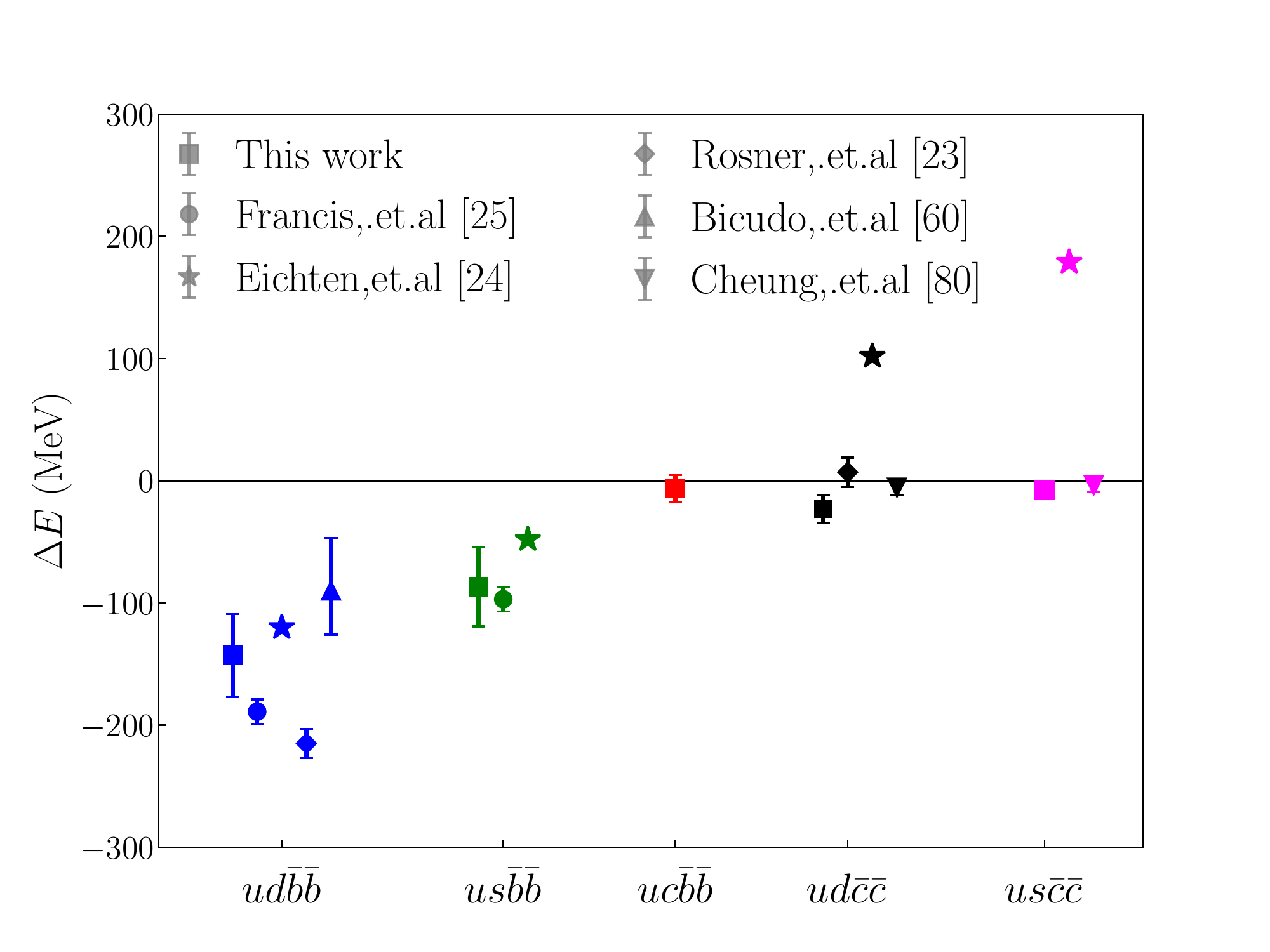}
	\caption{\label{fig:summary_spinone}Comparison of global results on the spin one doubly bottom and charm tetraquark states with various flavor combinations. $\Delta E$ is the energy difference between the ground state and the lowest strong decay threshold. Various flavor combinations represented on the horizontal axis are color coded as: blue, green, red, magenta and grey for the state $ud\bar{b}\bar{b}, \ us\bar{b}\bar{b}, \ uc\bar{b}\bar{b}, \ ud\bar{c}\bar{c}$ and $us\bar{c}\bar{c}$, respectively.}
\end{figure}

We also provide a comparison of global results of spin one doubly heavy tetraquark states with various flavors and show that in Figure~\ref{fig:summary_spinone}.
The results from Refs.~\cite{Eichten:2017ffp,Karliner:2017qjm} are based on HQET and potential model, respectively, while the rest are lattice calculations.
All results agree with the existence of deeply bound spin one tetraquark states, $ud\bar{b}\bar{b}$ and $us\bar{b}\bar{b}$, which are stable under strong interactions.
Our results for the doubly bottom states agree well with those from the HQET predictions \cite{Eichten:2017ffp} as well as that of the result in Ref~\cite{Francis:2016hui} at similar lattice spacings ($\sim 0.09$ fm). Ref~\cite{Francis:2016hui}  used $N_f = 2+1$  PACS-CS gauge field configurations and coulomb gauge fixed wall sources with clover action in the valence sector. The results were extracted at a single lattice spacing ($a \sim 0.09$ fm) at three pion masses and a chiral extrapolation with $m_{\pi}^2$ was performed to obtain the final result.
The result from Ref.~\cite{Bicudo:2015kna} were obtained from the potential based lattice QCD study where potentials of two $B$ mesons were computed in the static approximation for various spin-isospin combinations. These were then fitted to a phenomenologically motivated ansatz which were further used to solve a Schr\"odinger equation to determine a bound state. These calculations were performed at three pion masses ranging from $m_\pi \sim 340 - 650 $ MeV and the final results was obtained after chiral extrapolation. Ref. \cite{Cheung:2017tnt} used an anisotropic  $N_f = 2+1$ clover action and results were obtained at a single lattice spacing ($a_t \sim 0.0035$ fm with anisotropy 3.5)  and at a single pion mass ($m_{\pi} = 391$ MeV. For the doubly charm states, our results are in disagreement with those from the HQET results~\cite{Eichten:2017ffp}.
As we have showed earlier, this discrepancy is due to the deviation of HQET relations at the charm quark mass.


Inspired by the results in spin one sector, we also explore the spin zero tetraquark states with doubly bottom as well as with doubly charm quarks.
Here, we have computed flavor symmetric $uu\bar{b}\bar{b}$ and $uu\bar{c}\bar{c}$ states and also explored the pion mass dependence by dialing the light quark mass. 
To check the lattice spacing dependence of the observed results we perform the calculation on three different lattice spacings. 
In addition, we have also computed following flavor symmetric states, namely, $ss\bar{b}\bar{b}, ss\bar{c}\bar{c}$ and $cc\bar{b}\bar{b}$ at the physical strange, charm and bottom quark masses. 
For the doubly bottom state $uu\bar{b}\bar{b}$, we find that the energy splittings ($\Delta E^{0}$) are generally noisy and do not clearly exhibit a trend of increase in $\Delta E^{0}$ as the pion mass is lowered. 
Contrary to the results of its flavor antisymmetric cousin $ud\bar{b}\bar{b}$, the ground state energy of $uu\bar{b}\bar{b}$ coincides with its threshold at lower pion masses with no clear indication of any level below the threshold.
For the doubly charm state, $uu\bar{c}\bar{c}$, the extracted energy levels clearly lie above their respective thresholds with no discernible dependence on pion mass, again contrary to the results of its flavor antisymmetric cousin $ud\bar{c}\bar{c}$.
In performing the continuum extrapolation, no lattice spacing dependence is observed for $uu\bar{b}\bar{b}$ state while the $uu\bar{c}\bar{c}$ exhibits a mild dependence on the lattice spacing.
The flavor symmetric states $ss\bar{b}\bar{b}, ss\bar{c}\bar{c}$ and $cc\bar{b}\bar{b}$ exhibit similar qualitative features in that all the energy levels are found to be above their respective thresholds and no significant lattice spacing dependence is observed in the continuum extrapolation. 
Our final results for the spin zero sector are shown in Table \ref{tab:spinzero_final}
\begingroup
\renewcommand*{\arraystretch}{1.9}
\begin{table}[ht]
	\centering
	\caption{\label{tab:spinzero_final}{Final results for the spin zero tetraquarks}}
	\begin{tabular}{c c | c c  }\hline \hline 
		State &  $\Delta E^{\mathbf{0}}$ [MeV] & State &  $\Delta E^{\mathbf{0}}$ [MeV] \\ \hline \hline
		$uu\bar{b}\bar{b}$   & -5(18) & $uu\bar{c}\bar{c}$  & 26(11) \\ \hline
		$ss\bar{b}\bar{b}$  & 3(9)   & $ss\bar{c}\bar{c}$ & 14(4) \\ \hline
		$cc\bar{b}\bar{b}$   & 16(1) & & \\ \hline \hline
	\end{tabular}
\end{table}
\endgroup
In conclusion, the states in the spin zero sector do not indicate energy levels below their thresholds suggesting it is very unlikely that there exists any doubly heavy bound tetraquark state with spin zero.

The availability of energy values of spin one tetraquark states for a large number of light quark masses provide us an opportunity to investigate the mass relations (Eq.~(\ref{Eq:hqet_relation})) between different heavy flavored hadrons due to the heavy quark symmetry, as mentioned in Ref. \cite{Eichten:2017ffp}. 
For this, we redefine the relation as a ratio ($\mathcal{R}$) between different hadron masses (Eq.~(\ref{Eq:hqet_ratio})) where a value of unity justifies the validity of such mass relation, and any deviation
from unity indicates the amount of breaking of the heavy quark
symmetry at a given heavy quark mass. We find that for bottom quarks,
$\mathcal{R}_b = 0.837(38)$, indicating that the bottom quark is very
close to the heavy quark limit. On the contrary, at the charm quark
mass we find $\mathcal{R}_c = 0.602(22)$, which substantially deviates
from the heavy quark limit. This clearly suggests that the charm quark
is not heavy enough to impose heavy quark symmetry relations among
hadron masses such as in Eq.~(\ref{Eq:hqet_relation}), {\it i.e.}, as far those
mass relations are concerned one needs to be careful while treating
the charm quark within HQET.

The tetraquark states studied in this work are computed in a single volume.
 In order to make conclusive statements about their scattering amplitudes and complex poles, one needs to carry out similar studies on multiple volumes followed by a finite volume analysis \cite{Luscher:1990ck}. 
Such analysis will especially be useful for the states which are close to their thresholds. 
However, a comprehensive finite volume analysis for a calculation that is reported here requires significantly large computational resources. 
Currently that is beyond the scope of this work but we intend to pursue such finite volume analysis in the near future. However, it is worth noting that the finite volume corrections for many heavy tetraquarks, particularly for which $\Delta E$ values are large, will be substantially suppressed. This is because, as has been pointed out before \cite{Beane:2003da,Davoudi:2011md,Briceno:2013bda}, such corrections to the observed energy splitting are suppressed not only because of its large value but also for the large masses of the threshold states, which are two heavy mesons in these cases. It is thus expected that such tetraquark states will be stable under
strong interactions. Other errors related to our calculations, namely, unphysical sea quark mass, quark mass tuning, scale setting, mixed action effects, excited state contamination together will be much smaller compared to the statistical error \cite{Mathur:2018epb}, and the conclusion reached here will be unaffected by those. It will therefore be very useful to search
experimentally spin one doubly heavy tetraquarks particularly with two
bottom quarks, such as $ud\bar{b}\bar{b}$. However, it is very unlikely that there exists any doubly heavy bound tetraquark state with spin zero.



\begin{acknowledgments}
We are thankful to the MILC collaboration 
and in particular to S. Gottlieb for providing us with the HISQ lattices.  We like to thank R. V. Gavai for discussions, particularly on HQET relations and M. Hansen for the discussions on finite volume corrections. We also thank R. Lewis and M. Peardon for useful discussions.
Computations are carried out on the Cray-XC30 of ILGTI, TIFR,  
and on the Gaggle/Pride clusters of the Department of Theoretical Physics,
TIFR. P.J. and N. M. like to thank Ajay Salve, Kapil Ghadiali and P. M. Kulkarni for computational supports. M. P. acknowledges support from EU under grant no. MSCA-IF-EF-ST-744659 (XQCDBaryons) and the Deutsche Forschungsgemeinschaft under Grant No.SFB/TRR 55.

\end{acknowledgments}

\newpage
\bibliographystyle{utphys-noitalics}
\bibliography{heavy_tetraquarks}

\section{Appendix}
We tabulate the energy splittings, $\Delta E$, defined as the difference between the threshold energy and the ground state energy levels, of tetraquark states with various flavor-spin combinations as studied in this work.
\begingroup
\renewcommand*{\arraystretch}{1.4}
\setlength{\tabcolsep}{10pt}
\begin{table*}[ht]
	\centering
	\begin{tabular}{ccccccccc}\hline \hline
		$N^3_s \times N_t$ & $m_\pi$ (MeV) & $ud\bar{b}\bar{b}$ & $us\bar{b}\bar{b}$ & $uc\bar{b}\bar{b}$ & $ud\bar{c}\bar{c}$ & $us\bar{c}\bar{c}$ & $uu\bar{b}\bar{b}$ & $uu\bar{c}\bar{c}$ \\ \hline \hline
		$24^3 \times 64$                 & 689 & -83(9)    & -83(9)   & -15(4) & -11(3) & -11(3) & 17(10) & 11(2)\\ 
						 & 589 & -110(13)  & -101(9)  & -19(4) & -18(3) & -14(3) & 8(6)   & 9(2) \\
						 & 539 & -117(16)  & -104(9)  & -22(4) & -18(3) & -12(2) & 6(6)   & 11(2)\\ 
	 					 & 497 & -120(14)  & -100(14) & -18(6) & -22(5) & -13(3) & 8(8)   & 9(3) \\
	 					 & 449 & -127(18)  & -111(10) & -25(5) & -25(4) & -13(2) & 4(10)  & 10(3)\\
	 					 & 400 & -136(24)  & -111(12) & -21(5) & -27(5) & -12(3) & 0(11)  & 8(5) \\
	 					 & 367 & -145(21)  & -116(12) & -29(5) & -28(6) & -12(3) & -3(10) & 9(4) \\
	 					 & 337 & -146(25)  & -109(13) & -20(8) & -26(7) & -11(3) & -2(12) & 8(7) \\
	 					 & 297 & -164(36)  & -119(15) & -30(6) & -28(7) & -11(3) &   -    & 6(7) \\
	 					 & 257 & -181(43)  & -115(18) & -25(9) & -25(8) & -9(4)  &   -    &  - \\
	 					 & 237 &  -        & -112(21) & -29(8) &  -     & -      &   -    &  - \\
	 					 & 216 &  -        & -117(14) & -19(13)&  -     & -      &   -    &  - \\
	 					 & 202 &  -        & -126(18) & -27(11)&  -     & -      &   -    &  - \\
	 					 & 186 &  -        & -121(17) & -31(11)&  -     & -      &   -    &  - \\
	 					 & 153 &  -        &   -      & -33(13)&  -     & -      &   -    &  - \\ 
	 							    		\hline \hline
		$32^3 \times 96$                 & 688 & -62(13)  & -62(13) & -9(3)  & -13(3)  & -13(3) & 5(5)   & 12(3) \\
						 & 537 & -93(19)  & -77(15) & -12(5) & -19(5)  & -13(3) & -1(8)  & 9(7)  \\ 
                                                 & 491 & -123(25) & -74(23) & -14(5) & -23(6)  & -14(4) & -2(10) & 12(9) \\
						 & 441 & -135(21) & -79(18) & -12(5) & -23(8)  & -10(4) & -6(12) & 15(12)\\
						 & 396 & -147(31) & -91(17) & -16(5) & -27(10) & -10(5) & -6(13) & 19(8) \\
						 & 367 &  -       & -97(19) & -15(6) & -32(13) & -9(5)  & 0(11)  & - \\
						 & 345 &  -       &   -     & -17(6) &   -     & -      & -5(13) & - \\
		\hline \hline
		$48^3 \times 144$                 & 685 & -88(6)  & -88(6)  & -10(2) & -15(2) & -15(2) & 6(7) & 17(2)\\
						  & 645 & -94(7)  & -91(7)  & -11(2) & -15(2) & -13(3) & 4(7) & 17(3)\\
						  & 576 & -102(9) & -94(8)  & -10(2) & -15(3) & -13(2) & 3(8) & 18(4)\\
						  & 545 & -106(10)& -90(10) & -12(3) & -17(3) & -13(3) & -1(8)& 20(4)\\
		 \hline \hline
	\end{tabular}
	\caption{\label{tab:summary_splittings}{Summary of splittings of tetraquark states in this work.}}
\end{table*} 
\endgroup

\end{document}